\documentclass[3p,final,12pt,authoryear,bookmarks=false,colorlinks,linkcolor=blue,citecolor=blue,urlcolor=blue]{elsarticle}
\usepackage[utf8]{inputenc}
\usepackage{caption}

\usepackage{graphicx}
\usepackage{subcaption}
\usepackage{amsmath}
\usepackage{cite}
\usepackage{cancel} 
\usepackage{url}
\usepackage{setspace}
\usepackage{lscape}
\usepackage{lineno,hyperref}
\usepackage{amssymb}
\biboptions{authoryear}

\usepackage[figuresright]{rotating}
\begin{document}
\begin{frontmatter}



%\dochead{World Conference on Transport Research - WCTR 2023 Montreal 17-21 July 2023}%

\title{Exploring the combined effects of major fuel technologies, eco-routing, and eco-driving for sustainable traffic decarbonization in downtown Toronto}
%\title{Multi-objective anticipatory mixed-fuel green vehicle routing of connected and automated vehicles}


\author[a]{Saba Sabet} 
\author[a]{Bilal Farooq\corref{cor1}}

\address[a]{Laboratory of Innovation in Transportation (LiTrans), Toronto Metropolitan University, Toronto, M5B 1W7, Canada}
%\address[b]{Associate Professor, Transportation Engineering
%Canada Research Chair, Disruptive Transportation Technologies and Services Director, Laboratory of Innovations in Transportation (LiTrans), Toronto Metropolitan University, Toronto, M5B 1W7, Canada}

\textbf{Cite as: Sabet, S., Farooq, B. (2025) Exploring the combined effects of major fuel technologies, eco-routing, and eco-driving for sustainable traffic decarbonization in downtown Toronto, Transportation Research Part A: Policy and Practice, 192, 104385}

\begin{abstract}
{As global efforts to combat climate change intensify, transitioning to sustainable transportation is crucial. This study explores decarbonization strategies for urban traffic in downtown Toronto through microsimulation, evaluating the environmental and economic impacts of vehicle technologies, traffic management strategies (eco-routing), and driving behaviors (eco-driving). The study analyzes 140 decarbonization scenarios involving different fuel types, Connected and Automated Vehicle (CAV) penetration rates, and anticipatory routing strategies. Using transformer-based prediction models, we forecast Greenhouse Gas (GHG) and Nitrogen Oxides (NOx) emissions, along with average speed and travel time.}
{The key findings show that 100\% Battery Electric Vehicles (BEVs) reduce GHG emissions by 75\%, but face challenges related to cost and infrastructure. Hybrid Electric Vehicles (HEVs) achieve GHG reductions of 35-40\%, while e-fuels result in modest reductions of 5\%. Integrating CAVs with anticipatory routing strategies significantly reduces GHG emissions. Additionally, eco-driving practices and eco-routing strategies have a notable impact on NOx emissions and travel time.}
{By incorporating a comprehensive cost analysis, the study offers valuable insights into the economic feasibility of these strategies. The findings provide practical guidance for policymakers and stakeholders in developing effective decarbonization policies and supporting sustainable transportation systems.}

\end{abstract}

\cortext[cor1]{Corresponding author. bilal.farooq@torontomu.ca Tel.: +1-416-979-5000.}
\end{frontmatter}

%\correspondingauthor[*]{Corresponding author. Tel.: +0-000-000-0000 ; fax: +0-000-000-0000.}
%\email{saba.sabet@torontomu.ca; bilal.farooq@torontomu.ca}

% \linenumbers

\vspace*{-12pt}
%\enlargethispage{-7mm}
\section{Introduction}
\label{main}

The transportation sector is one of the few remaining sectors that still heavily rely on fossil fuels, leading to significant Greenhouse Gas (GHG) emissions, which contribute to global warming and environmental challenges. In 2017, the transportation sector ranked as the second-largest emitter of GHGs, responsible for a substantial 174.4 million metric tons of carbon dioxide equivalent (MT CO2e) \citep{United}. To address this pressing issue, stringent limitations have been imposed on emissions from transportation industries, necessitating a transition towards more sustainable practices. Among transportation sources, passenger cars account for nearly 41\% of global transportation emissions, making them a key target for emission reduction efforts \citep{jenn2019alternative}.  There is an urgent need for reducing this carbon footprint through emerging technologies, policies, and behavioural changes. 
Transportation decarbonization refers to efforts to reduce the carbon emissions associated with the transportation sector. This can include a variety of technologies such as alternative fuels, connected and automated vehicles, and machine/deep/reinforcement learning; strategies such as system optimal routing; and promoting behavioural change in travellers such as encouraging driving behaviour that results in less emissions from the vehicles. The goal of transportation decarbonization is to reduce greenhouse gas emissions and mitigate the impacts of climate change. In another observation of emission reduction research, \citet{bhardwaj2020have} develops a framework for examining policy interactions across criteria like GHG mitigation, cost-effectiveness, political acceptability, and transformative signal. \citet{tzeiranaki2023impact} delves into historical energy consumption trends within the EU road transport sector, examining the role of key determinants like economic growth, population growth, fuel prices, and vehicle fleet characteristics.

To cut energy use and emissions, we're focusing on alternative energy sources as the primary post-fossil-fuel power. Still, these alternatives might not always offer big benefits when considering their full impact \citep{mccubbin1996social}.  Many countries are also pushing for zero-emission vehicles (ZEVs) soon. Alongside this, emerging tech like connected and automated vehicles could quickly impact emissions, mobility, and fuel use. With regards to the technological aspect, Battery Electric Vehicles (BEVs) are a prominent solution in the context of electrification, offering an immense potential to substantially reduce GHG emissions compared to conventional Internal combustion Engine Vehicles (ICEVs). Additionally, BEVs bring about supplementary benefits such as lower energy costs and decreased maintenance requirements. On the other hand, HEVs provide an intermediate step by combining electric power with an internal combustion engine, reducing fuel consumption and emissions to a great extent. In the quest to further reduce GHG and Nitrogen Oxides (NOx) emissions using technology, e-fuels \citep{ueckerdt2021potential}, also known as synthetic fuels, are produced using renewable energy sources and can be used in conventional ICEVs without major modifications. These fuels offer the potential to reduce the carbon footprint in the transportation sector while utilizing existing infrastructure and vehicle fleets.
However, the optimal combination of e-fuels, BEVs, HEVs, and ICEVs to achieve the greatest reduction in GHG and NOx emissions remains an area of debate. To comprehensively evaluate the environmental impact, it is essential to consider factors such as vehicle types, their market shares, and the associated emissions across their lifecycle stages. The need to assess the cost, availability, travel time, and emissions compromises underpins the pivotal question of whether a singular approach or a combination of strategies is most effective for successful transportation decarbonization. In light of the pressing need for transportation decarbonization, the pivotal questions that demand exploration are whether a singular approach stands out as superior or if a harmonious combination of technologies and strategies is the optimal path forward. Understanding the respective pros and cons of electrification, adoption of e-fuels, Hybrid Electric Vehicles (HEVs), and eco-routing and eco-driving techniques will be crucial in determining their effectiveness and potential limitations and to answer the key question of which pathways are more crucial to follow when considering users, manufacturers, and policymakers benefits. 

This research aims to assess the potential for decarbonization in urban traffic, using technological, policy, and behavioural pathways by microsimulating 140 scenarios. Specifically, we consider the passenger vehicle fleet operating in downtown Toronto that contributes significantly to greenhouse gas (GHG) and Nitrogen Oxides (NOx) emissions. We acknowledge the possibility that any significant mode shift may not be achievable in the foreseeable future, so decarbonization options related to passenger vehicles must be explored. Our comprehensive examination revolves around four major pathways: electric vehicles (EVs), hybrid electric vehicles (HEVs), e-fuels, and the integration of eco-efficient technologies.
Our study also evaluates the role of Connected and Automated Vehicles (CAVs) and their influence on emission levels under different penetration scenarios (0\%, 50\%, and 100\%). We investigate how CAVs, by optimizing routing, reducing idling, and improving traffic flow, can further contribute to GHG emissions reduction.
Furthermore, the integration of eco-driving and eco-routing practices is investigated. Eco-driving, focusing on optimizing driving behaviour, and eco-routing, aiming to identify the most fuel-efficient routes, are explored for their potential synergistic effects on emission reduction.
The findings will provide valuable insights into the development of sustainable transport policies and practices.

The research objectives of this study are to answer the questions regarding the potential of combining emerging vehicle technologies, traffic management, and driving behaviour change:
\begin{enumerate}
    \item Which pathways have the most effect on mitigating adverse effects of NOx and GHG while considering commuting and other costs?
    \item How much can CAVs, eco-routing and eco-driving adoption reduce GHG and NOx in a mixed-fuel environment?
    \item What combinations and penetration rates of vehicles with different fuel types, are the most desirable for GHG and NOx reduction, if a full life-cycle analysis from well to wheel is to be considered for emission cost?
\end{enumerate}

The existing literature primarily tends to focus on individual aspects of decarbonization, neglecting the comprehensive evaluation of different pathway combinations, which are likely to yield more realistic and impactful results. This research aims to bridge this critical gap by thoroughly analyzing the synergistic effects of various technologies, routing strategies, driving behaviour, and market penetrations to identify the most effective and practical approach for significant emissions reduction in the transportation sector. In this study,  we systematically examined Well-To-Wheels (WTW) GHG and NOx emissions from vehicles with different types of fuel based on the criteria and numbers from the 2021 report of the United States Department of Energy \citep{kelly2022cradle}. Well-to-Wheels (WTW) refers to the comprehensive assessment of greenhouse gas (GHG) and nitrogen oxides (NOx) emissions throughout the entire lifecycle of vehicles, encompassing all stages from fuel production to vehicle operation.,
We assess the interplay between vehicle types, routing strategies, and market shares to determine the most effective approach for mitigating climate change and promoting sustainable transportation practices.
Our research thus offers a holistic assessment of decarbonization strategies in the transportation sector with a very high-resolution microsimulation, aiming to inform policy decisions, support sustainable urban planning, and contribute to global climate change mitigation efforts. The insights gathered through this comprehensive evaluation can inform policy decisions, urban planning, and transportation management strategies, paving the way for a greener and more sustainable transportation system. 

The rest of the paper is organized as follows; in Section \ref{litrev}, literature review is discussed, the methodology utilized in the current study is in Section \ref{meth}, the results of our study are presented in Section \ref{res}, planning and policy implications are presented in Section \ref{pol}, limitations in Section \ref{limits}, and the concluding discussion and future direction is in Section \ref{con}. 

\section{Literature Review}
\label{litrev}
This section presents an overview of the recent literature on the topics explored in this study. The overview of the background literature is divided into different parts based on the pathways explored in the current study. 
\subsection{Alternative fuels}
As a technological solution, alternative fuels play a crucial role in reducing GHG and NOx emissions, offering a more sustainable and environmentally friendly option compared to conventional fossil fuels. They contribute to mitigating climate change, improving air quality, and reducing dependence on fossil fuel resources. Implementing alternative fuels in the transportation network is a key aspect of strategic planning for transportation decarbonization. Major alternative fuels may include electricity, hydrogen, biofuels, e-fuels and natural gas. These fuels can significantly reduce greenhouse gas emissions compared to traditional fossil fuels such as gasoline and diesel \citep{rony2023alternative}.

Considering the existing body of research on alternative fuels' role in reducing transportation-related carbon emissions, electric vehicles has gained prominence over the last decade. This is due to their potential as a viable alternative to traditional internal combustion engines \citep{diaz2020electric}. The prevailing literature generally suggests that battery electric vehicles (BEVs) tend to have lower life cycle greenhouse gas (GHG) emissions and consume less energy compared to internal combustion engine vehicles (ICEVs) in most scenarios \citep{wu2018life}. To elaborate further, the production phases of BEVs contribute more to GHG emissions due to the sourcing and processing of raw materials, as well as the vehicle manufacturing process, when contrasted with ICEVs. However, this disparity is balanced by lower emissions during the usage phase of the vehicle, contingent on the source of electricity utilized to charge the BEV batteries. Consequently, the significance of the electricity generation source for battery charging is a crucial factor. 
E-fuels, also known as synthetic or renewable fuels, are produced using renewable energy sources. These fuels can be used in conventional internal combustion engines and have the potential to contribute to GHG and NOx reduction. The production process often involves capturing and recycling CO2 emissions, making them a potential pathway for reducing overall emissions in sectors that rely on liquid fuels. The use of certain e-fuels, particularly advanced biofuels, can result in lower NOx emissions compared to conventional fossil fuels. Careful production and refining processes can help minimize emissions of NOx and other pollutants associated with combustion \citep{ueckerdt2021potential}. HEVs and e-fuels play transitional roles by reducing emissions during the transition to a fully electrified vehicle fleet. Additionally, e-fuels can provide a sustainable alternative for sectors where electrification may be challenging, such as aviation and heavy-duty transportation.

The Fraunhofer Institute's perspective on the potential role of synthetic fuels and hydrogen-based drive technologies in future transportation raises intriguing possibilities \citep{jensterle2019role}. While they acknowledge the potential insignificance of these technologies in private vehicles, they highlight their significance in long-distance for last mile delivery. However, it is crucial not to overlook the negative outcomes of hydrogen-based fuel cell technology, which poses high efficiency and operating cost challenges. The study by \citet{horvath2018techno} compares battery-powered and hydrogen-powered electric vehicles, revealing that the latter suffers from lower efficiency rates due to energy losses during hydrogen production and conversion. 
Despite the potential prevalence of alternative fuels in the long term, it seems improbable that they will completely replace fossil fuels in the near future. Instead, the focus should be on prioritizing alternative fuels for transportation sectors where they are cost-effective and have a competitive advantage, considering their limited availability. Moreover, the complexities and additional infrastructure required for handling hydrogen make its adoption challenging, particularly for heavy-duty transportation, which necessitates a hydrogen grid and fuel cells. Given these considerations, it is evident that studies on eco-routing are more likely to be influenced by research on Electric Vehicles (EVs) and Hybrid Electric Vehicles (HEVs) in the foreseeable future, as the penetration of alternative fuel vehicles is currently scarce. To delve into AFV-related aspects of decarbonization, a sufficient presence of such vehicles on the transportation network is required, and the research in this area remains limited. Notably, scholars tend to focus on AFV studies in urban transit, logistics, air, and rail transportation, where the use of alternative fuels, especially in heavy-duty vehicles, is more favourable. As demand and penetration rates of alternative fuel vehicles increase, it is anticipated that AFV studies will experience substantial growth in the future.

{
While the transition to electrification, especially for personal vehicles, is gaining traction, the role of alternative fuels beyond electricity, particularly for heavy-duty transportation and aviation, is not as extensively explored. For example, it has been pointed out that E-fuels have the potential to be sustainable alternatives for these harder-to-electrify sectors, but more research is needed to assess their real-world performance and long-term environmental and economic viability. In addition, the impact of the electricity generation source used to charge the vehicle batteries needs to be investigated in detail, as it can significantly influence the life cycle GHG emissions of electric vehicles.
}

\subsection{Connected and Automated Vehicles}
The National Highway Traffic Safety Administration (NHTSA) \citep{thierer2016comment} defines varying levels of vehicle automation. Level 2 or 3 automation vehicles can help mitigate human errors and offer potential benefits. Connected and Automated Vehicles (CAVs) can significantly increase energy efficiency due to their advanced sensing, processing, and control abilities. Earlier studies primarily focused on CAV software and safety, leaving energy efficiency overlooked. CAVs can improve road safety, efficiency, and even affect nearby traffic's energy efficiency due to their precise control. They enable cooperation via vehicle-to-infrastructure links, enhancing traffic flow, safety, and lowering energy use. While energy efficiency wasn't a central CAV development aspect, it carries substantial weight.  
\citet{wadud2016help} conducted a scenario analysis suggesting vehicle automation could cut energy use and greenhouse gas emissions in an optimistic scenario by half, yet double them under different circumstances. Various analyses stress the outcomes' dependence on dominant scenarios, calling for proactive policy to guide tech toward energy efficiency \citep{alexander2016impact}. In addition, in another study, \citet{greenblatt2015automated} pointed out potential energy and environmental benefits while cautioning about automation's game changing impact. \citet{rahman2023impacts} recently systematically reviewed AVs' effects on urban transport and the environment.
Few scholars, including \citet{alfaseeh2019multi} and also  \citet{djavadian2020multi}, have explored novel variants in eco-routing literature, addressing connectivity and automation issues. 
Their objective was to reduce travel time while cutting down on greenhouse gas (GHG) and nitrogen oxide (NOx) emissions. They employed diverse costing methods and routing tactics for this purpose. Additionally, their research underscores the significance of looking into the effects of Vehicular Ad hoc Networks (VANETs) on eco-routing, particularly within the context of Intelligent Transportation Systems (ITS) applications. They took into account factors like the availability of network data, its reliability, precision, as well as its temporal and spatial distribution.
To delve deeper, readers can refer to \citet{RePEc:spr:sprchp:978-3-030-97940-9_126} in which the same authors of the current study have investigated energy smart transportation systems with the focus on CAVs, and AFVs.

{While the potential energy efficiency benefits of CAVs have been highlighted, many of the previous studies focus primarily on their software intelligence and safety aspects. The energy efficiency of CAVs and their potential for reducing emissions remain largely unexplored, with the impact on energy use not being a main part of the development of CAVs. This is an essential area for further research, as the future of transport is expected to feature higher levels of automation. Also, the limited exploration of the impact of Vehicular Ad hoc Networks (VANETs) on eco-routing within the context of Intelligent Transportation Systems (ITS) applications, and the lack of comprehensive investigation into factors such as network data availability, robustness, accuracy, and temporal and spatial distribution are considered the potential gaps in the literature in this context.}

\subsection{Eco-routing}
Eco-routing or green-routing is a specific tactic that can be used in the tactical planning of decarbonization in the transportation sector. It involves using advanced algorithms and data analysis to optimize the routes of vehicles and reduce the carbon emissions associated with transportation.
This can include a variety of different strategies, such as: optimizing routes to reduce fuel consumption and GHG emission, avoiding congested areas to reduce emissions from idling, inventory routing of vehicles to take advantage of electric charging infrastructure where appropriate green-routing can also take into account other factors such as reducing travel time, considering traffic patterns, weather conditions, and road closures, to ensure the most efficient and sustainable routes are taken \citep{alfaseeh2019multi}. 
By using green-routing, organizations can decrease their carbon footprint and reduce their environmental impact while improving their operational efficiency. It also can be used in logistics, public transportation and delivery services \citep{rezaei2019green}. 
In their comprehensive evaluation of existing eco-routing research, \citet{alfaseeh2020deep} found that previous exploratory studies on the subject relied heavily on macroscopic traffic and emission models, with a focus on small-scale case studies. These studies typically employed centralized routing systems and are targeted at optimizing a single routing objective at a time.

In an effort to address these shortcomings, \citet{alfaseeh2019multi} implemented pro-active multi-objective eco-routing within a distributed routing framework. Their methodology utilized a per lane weighted average for assessing GHG costs on various routes. They found that while using this approach resulted in an underestimation for routes with a larger number of lanes, there is a noticeable reduction in travel time and emissions when applying multi-objective routing.

{Existing eco-routing studies mostly focus on operational level routing decisions, with less emphasis on other supply chain management aspects, such as network design, road tolls, and reliability index. Additionally, the topic of the Alternative Fuel Vehicle Routing Problem (AFVRP) in mixed fleets and with respect to connected and autonomous vehicles remains underexplored. The demand side of AFVRP based on fuel technologies also requires further development.
}
\subsection{Eco-driving}
Eco-driving, also known as fuel-efficient driving, is an example of operational level planning that can help drivers change their driving behaviour and reduce fuel consumption and emissions. Some of the key principles of eco-driving include \citep{barth2011dynamic}:
\begin{itemize}
    \item \emph{Smooth acceleration and deceleration:} Avoiding sudden accelerations and decelerations can help to reduce fuel consumption and emissions.
    \item Anticipating traffic: By anticipating traffic flow and potential hazards, drivers can reduce the need for sudden braking and acceleration.
    \item \emph{Proper vehicle maintenance:} regularly maintaining a vehicle, including keeping tires properly inflated and ensuring that the engine is running efficiently, can help to reduce fuel consumption.
    \item \emph{Avoiding idling:} Turning off the engine when a vehicle is not in motion can help to reduce fuel consumption and emissions.
\end{itemize}
By adopting eco-driving techniques, drivers can reduce their own fuel consumption and emissions, and also help to reduce traffic congestion and improve air quality. Eco-driving can be incorporated into operational level planning in transportation decarbonization as a way to improve the efficiency of the transportation system and reduce emissions. Eco-driving can be supported by advanced vehicle technologies, such as hybrid and electric vehicles, and connected and autonomous vehicles which can adapt to traffic conditions and optimize energy use.

Compared to research on energy perception in HEVs and BEVs, fewer studies have explored driving control for energy management. This discrepancy stems from the intricate relationship between energy consumption and vehicle dynamics. While eco-driving primarily targets car, truck, and bus drivers, it also extends its benefits to railway, tractor, and construction machinery operators. A day of eco-driving training can lead to fuel reductions of 10\% to 20\%, depending on the driver's current habits. This positions eco-driving as a potent strategy for cutting greenhouse gas emissions, surpassing even vehicle replacement. Another advantage is the rapid cost recovery, taking less than a year for individual drivers.
Eco-driving yields multiple advantages such as smoother traffic, reduced stress, enhanced safety, and lowered operating costs and emissions. These gains can be achieved within the same or shorter travel time due to the adaptable nature of eco-driving. Recent data indicates that eco-drivers can reach destinations faster due to improved traffic flow and fewer stops. While applicable to older vehicles, its full potential is harnessed with modern technologies, including advanced engines \citep{shen2020minimum}.
However, the universality of eco-driving is limited. It can exacerbate congestion on busy roads but proves advantageous in less crowded areas. Larger cities with higher congestion levels require customized adaptations. Effective measures include speed restrictions on high-capacity roads and infrastructure enhancements like traffic lights, bus lanes, and dual carriageways for areas with lower congestion \citep{coloma2020developing}.
Despite encouraging findings, eco-driving research faces limitations. Addressing these challenges could involve integrating Internet of Vehicles (IoV) and cloud computing platforms. Existing studies primarily focus on individual vehicle speed planning for fuel efficiency, neglecting fleet or network-level energy consumption and emissions \citep{xu2021overview}.

{In a more recent study, \citet{naeem2023energy} studied energy economizing using Connectivity-based eco-Routing and Driving (CeRD) for a fleet of battery electric vehicles (BEVs) which considered three pathways out of four, discussed in our study. Their study was based on advising BEVs on energy-efficient routes by considering factors such as traffic conditions, vehicle load, speed limits, and battery constraints.} 

{Eco-driving holds significant potential for reducing GHG emissions. However, the complex interactions between energy consumption and vehicle dynamic characteristics pose a challenge for research in this area. More studies are needed that focus on the different variables of eco-driving and how they interact with each other. In addition, while eco-driving techniques can be used with older vehicles, understanding how they can be best applied with advanced vehicle technologies is crucial. This could potentially include the use of AI and Machine Learning techniques to optimize eco-driving strategies.
}

\subsection{Concluding Remarks}
{Based on the comprehensive overview of existing literature on various aspects of transportation decarbonization, it's apparent that several areas warrant further investigation. Table \ref{table:summary} shows a summary of the studies mentioned in the literature review.}
As displayed in the table, no study has covered all the pathways before. Therefore, addressing these research gaps is of critical importance to develop more comprehensive strategies for transportation decarbonization. Each of these areas represents a pathway toward reducing GHG emissions in the transportation sector, and understanding them more fully will enable the development of more effective policies, technologies, and practices for a sustainable transport future. Therefore, future research should focus on these areas to fill the existing knowledge gaps and drive the transition toward sustainable and decarbonized transportation.

\newpage
\begin{table}[!ht]
\centering
\resizebox{\textwidth}{!}{%
\begin{tabular}{|p{3cm}|p{2.85cm}|p{2.5cm}|c|c|p{5cm}|}
\hline
\textbf{Case Study} & \textbf{Alternative Fuels Comparision} & \textbf{CAVs} & \textbf{Eco-routing} & \textbf{Eco-driving} & \textbf{Notes} \\ \hline
Rony et al., 2023 & \multicolumn{1}{|c|}{\checkmark} & & & & Focus on the role of alternative fuels in reducing GHG emissions \\ \hline
Diaz, 2020 & \multicolumn{1}{|c|}{\checkmark} & & & & Focus on electric vehicles as an alternative to ICEVs \\ \hline
Wu et al., 2018 & \multicolumn{1}{|c|}{\checkmark} & & & & Comparison of GHG emissions between BEVs and ICEVs \\ \hline
Ueckerdt et al., 2021 & \multicolumn{1}{|c|}{\checkmark} & & & & Analysis of e-fuels and their potential for reducing NOx emissions \\ \hline
Horvath et al., 2018 & \multicolumn{1}{|c|}{\checkmark} & & & & Comparison between battery-powered and hydrogen-powered EVs \\ \hline
Jensterle et al., 2019 & \multicolumn{1}{|c|}{\checkmark} & & & & Discussion on the future role of synthetic fuels and hydrogen-based drive technologies \\ \hline
Thierer and Watney, 2016 & & \multicolumn{1}{|c|}{\checkmark} & & & Overview of levels of vehicle automation \\ \hline
Wadud et al., 2016 & & \multicolumn{1}{|c|}{\checkmark} & & & Scenario analysis on energy efficiency of automated vehicles \\ \hline
Alexander-Kearns et al., 2016 & & \multicolumn{1}{|c|}{\checkmark} & & & Stressing the need for policies guiding tech toward energy efficiency \\ \hline
Greenblatt and Shaheen, 2015 & & \multicolumn{1}{|c|}{\checkmark} & & & Potential energy and environmental benefits of vehicle automation \\ \hline
Rahman and Thill, 2023 & & \multicolumn{1}{|c|}{\checkmark} & & & Review of AVs’ effects on urban transport and the environment \\ \hline
Alfaseeh et al., 2019 & & \multicolumn{1}{|c|}{\checkmark} & \checkmark & & Exploring eco-routing and connectivity issues \\ \hline
Djavadian et al., 2020 & & \multicolumn{1}{|c|}{\checkmark} & \checkmark & & Focus on eco-routing and automation \\ \hline
Sabet and Farooq, 2023 & \multicolumn{1}{|c|}{\checkmark} & \multicolumn{1}{|c|}{\checkmark} & & & Investigation into energy smart transportation systems \\ \hline
Alfaseeh and Farooq, 2020 & & & \checkmark & & Evaluation of existing eco-routing research \\ \hline
Rezaei et al., 2019 & & & \checkmark & & Application of green-routing in logistics and public transportation \\ \hline
Barth et al., 2011 & & & & \checkmark & Key principles of eco-driving were developed\\ \hline
Shen et al., 2020 & & & & \checkmark & Impact of eco-driving on traffic flow and travel time \\ \hline
Coloma et al., 2020 & & & & \checkmark & Customized eco-driving strategies for different city congestion levels \\ \hline
Xu et al., 2021 & & & & \checkmark & Challenges in integrating IoV and cloud computing with eco-driving \\ \hline
Naeem et al., 2023 &  & \multicolumn{1}{|c|}{\checkmark} & \checkmark & \checkmark & Energy optimization using Connectivity-based eco-Routing and Driving (CeRD) for BEVs \\ \hline
Our work &  \multicolumn{1}{|c|}{\checkmark} & \multicolumn{1}{|c|}{\checkmark} & \checkmark & \checkmark & Deatiled microsimulation on a relatively large network \\ \hline
\end{tabular}%
}
\caption{{Summary of the major dimensions studied in existing literature}}
\label{table:summary}
\end{table}

\section{Methodology}
\label{meth}
%When considering the overall emissions associated with different fuel types, it is essential to evaluate the ``well-to-wheel'' emissions. This term encompasses the entire life cycle of the fuel, including its production, distribution, and use in vehicles.

\begin{figure}
     \includegraphics[width=\textwidth]{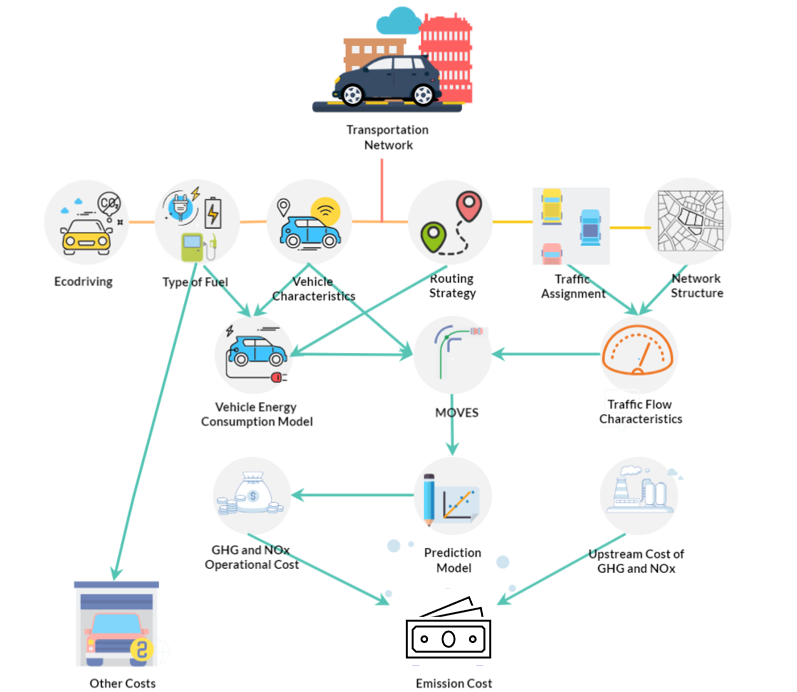}
     \caption{The proposed methodology of the current study}
     \label{fig:z}
 \end{figure}

The flowchart in Figure \ref{fig:z} describes the methodology followed in this study. %Information about the network structure and traffic assignment have already been covered in the case study section. 
It elaborates on the specifications of the traffic and emission models, GHG and NOx costing approaches, adopted emission rate and speed predictive models, eco-driving model and the routing strategy considered. The estimated second-by-second speed and acceleration are used for calculating emission parameters.
After defining the best emission costing approach, different scenarios, which considered 0, 25\%, 50\%, 75\%, and 100\% of BEV, HEV, and e-fuel vehicle penetration rates, are investigated in connected and automated vehicles. Then the same procedures are applied to the semi- and non-CAV environments with mixed-fuel vehicles. Eco-driving has also been investigated in accordance with different routing strategies including user equilibrium, myopic and anticipatory routing. A total of 140 scenarios are investigated in this study based on different variations of CAV, fuel type, eco-driving and routing strategies. The selection of penetration rates at 0\%, 25\%, 50\%, 75\%, and 100\% allows for a comprehensive analysis of how different levels of adoption impact emissions. Additionally, the research considers semi- and non-CAV environments with mixed-fuel vehicles to assess how different vehicle technologies and automation levels influence emissions.

Furthermore, the investigation of eco-driving in accordance with different routing strategies, such as user equilibrium, myopic, and anticipatory routing, adds depth to the analysis. This exploration allows for the understanding of how routing decisions can influence energy consumption and emissions, contributing to a more holistic approach in evaluating the potential of eco-routing strategies. Finally, a comprehensive comparison has been conducted between the scenarios to illustrate the impact of different EVs penetration rates, eco-driving and eco-routing implementation, and connectivity and automation, both at the operational level (tailpipe emission) and from well-to-wheel (upstream + tailpipe emission).

\subsection{Network Structure}
%The network structure, as described in the case study section includes downtown Toronto's 839 links which are the streets and 268 nodes which are intersections.
%The E2ECAV system is deployed on the dense urban network (see Figure \ref{dt}. This proposed system is implemented in an agent-based traffic simulation programmed in MATLAB.
%The time dependent exogenous demand Origin–Destination (OD) matrices are based on five-minute intervals from TTS. 
While the vehicle movement is simulated every second, the link level space mean speed, density, and flow information are collected in one-minute intervals as it is found to be the optimal updating interval based on \citet{alfaseeh2018impact}. The availability of link-level space mean speed monitoring is assumed to supply dynamic travel time-based shortest paths. This assumption is based on the extensive sensor infrastructure already in place in downtown Toronto, capable of providing a near real-time state of the network. 
Three different agents and two network layers create the simulated environment. In this regard, the network layers are contained the communication network and the physical road network. To be more specific, $G(I,L)$ shows the physical road network, in which $I$ is the sign of intersections and $L$ represents links. In addition, the three agents in this system are defined as follows: vehicle agents ($v \in V$), passenger agents ($p \in $P), and infrastructure agents. Vehicle agents also would be divided into two main types, including normal traffic ($M$) and CAV fleet ($f \in $F). It has to be mentioned that both of these types can be either electric or fossil fuel vehicles. Infrastructure agents consist of two link agents, ($l \in $L) and intelligent intersection agents ($I_n^2 \in I^2$). Finally, communications contain two major types of $(V2I)$ and $(I2I)$. Furthermore, $\Delta$ is the dispatch update cycle which is considered one minute in this study. 

%In this study, the second-by-second traffic and also environmental variables are gathered to find out the space mean link indicators, including speed (km/hour), density (vehicle/km), flow (vehicle/hour), and GHG Emission Rate (ER) (gram), and NOx (gram). Accordingly, the simulation finished when all vehicles arrived at their destinations. During the simulation, the amounts of speed, density, GHG, NOx, and flow are updated every minute. In terms of emission modeling, the Motor Vehicle Emission Simulator (MOVES), developed by the United States Environmental Protection Agency (USEPA), is employed to generate GHG ERs (in CO2eq g/s) \citep{wang2020movestar}. The MOVES model estimates second-by-second emissions based on the vehicle operating mode, which is determined by the Vehicle-Specific Power (VSP) as defined in Eq. (6). The estimated second-by-second vehicle speed and acceleration by the IDM model are used as inputs for MOVES to estimate the vehicular second-by-second GHG emissions. This method of GHG estimation provides a more accurate reflection of the environmental state \citep{prevention2020united}, making it suitable for routing applications. The vehicle composition data relied on the report by the Ministry of Transportation Ontario \citep{alfaseeh2020deep}. The simulations in this study are executed using three computers equipped with Core i7 CPUs, 64-bit versions of the Windows 11 operating system, and 32.0 GB RAM.

\subsection{Traffic Assignment}
Using data from the Transportation Tomorrow Survey (TTS), a travel log survey in the Greater Toronto and Hamilton Area, this study focuses on evaluating the impact of connected and automated controls on network supply. Notably, the study intentionally excludes any demand changes stemming from capacity alterations, concentrating solely on network throughput effects on GHG emissions and air quality. As a result, any ``induced" demand arising from changes in network capacity is intentionally excluded from the study's scope. The simulation uses a one-second time step and one-minute travel time updates. To address variability, five replications with distinct random seeds are performed for each of the two scenarios, with the results averaged. Each scenario concludes when all vehicles reach their destinations.

\subsection{Routing Strategies}
The routing strategies used in this study include (a) assigning the minimum travel time route based on the traffic conditions at the start of the trip, resulting in a user equilibrium assignment, and (b) assigning and updating real-time routes using E2ECAV framework with myopic and anticipatory routing, resulting in an approximate system optimal routing (for evidence, see \citet{djavadian2020multi}). User equilibrium state assumes that each driver acts independently and optimally to minimize their individual travel time, leading to a state of equilibrium in which no driver can decrease their travel time by unilaterally changing routes. Myopic routing is based on the principle of short-term decision making, focusing primarily on the immediate conditions of the network. 
On the other hand, anticipatory routing takes a more forward-thinking approach, incorporating future traffic and environmental conditions into its decision-making process. Anticipatory green routing takes the emission prediction of downstream links in future time-step, to decide and provide a dynamic route for vehicles. In the previous studies, it is found that anticipatory routing outperformed the myopic strategy regardless of the routing objective \citep{alfaseeh2020deep}. In this study, the effects of myopic and anticipatory routing strategies, integrated with other technologies have been examined for a comparative analysis.

\subsection{Types of Fuel}

In our methodology, we have chosen to focus on e-fuels and fossil fuels for conventional internal combustion vehicles and for electric vehicles, Battery Electric Vehicles (BEVs), and Hybrid Electric Vehicles (HEVs). 0\%, 25\%, 50\%, 75\% and 100\% combinations of the four fuel types have been considered to generate the scenarios. %This selection is crucial as it represents both the current dominant fuel types and the emerging sustainable alternatives.
E-fuels are important due to their potential as a carbon-neutral energy source. Fossil fuels, although environmentally problematic, remain the primary energy source for transportation worldwide, making their consideration vital. BEVs and HEVs are at the forefront of the transition towards sustainable transportation, with BEVs offering zero-emission transport and HEVs combining the benefits of combustion engines and electric motors for enhanced efficiency. These four categories allow us to comprehensively evaluate the current and future directions of the transportation sector.

\subsection{Eco-driving}
Eco-driving is investigated in this study to assess its potential impact on energy consumption and emissions in the transportation sector, as it involves optimizing driving behavior to achieve fuel efficiency and reduce greenhouse gas (GHG) and nitrogen oxides (NOx) emissions. By investigating eco-driving in combination with different routing strategies and fuel types, the study aims to identify effective approaches for promoting sustainable and environmentally friendly transportation practices. In this regard, in the current study, all scenarios of different CAV MPRs, fuel types and routing strategies are investigated with and without including eco-driving techniques. In this regard, the eco-driving is implemented in the simulation using the method proposed by  \citet{zavalko2018applying}. Zavalko's methodology involved using an established in-vehicle device to track various acceleration levels during vehicular operation at consistent, brief intervals. This device, supplemented by its dedicated software, determined comprehensive energy indices: acceleration energy, energy dissipated during braking, steady movement energy, and the cumulative energy supplied to the driving wheels, all based on speedometer readings.
During each of these intervals $(t_i)$, the device collected data on the distance covered $(S_i)$ from the speedometer, which is used to calculate velocity $(V_i)$ and acceleration $(j_i)$ through specific equations. 
%$$
\begin{align}
V_i &=\displaystyle \frac{S_i}{t_i}(\text{m/s}) \\
j_i &=\displaystyle \frac{V_{i-1}-V_i}{t_i}(\text{m/s}^2)
\end{align}
%$$

Based on the acceleration $j_i$ and velocity $V_i$ values, the time period is classified as acceleration $\left(j_i>0, V_i>0\right)$, deceleration $\left(j_i>0, V_i>0\right)$, steady movement $\left(j_i=0, V_i>0\right)$ or stop $\left(j_i=0, V_i=0\right)$. The software then calculated the energy indices for each phase. Taking into account certain simplifications, the formulas for computing energy indices are adapted. For instance, the software calculated braking energy $(E_b)$ using specific formulas that accounted for factors such as the rotational-inertia coefficient, vehicle mass, and average acceleration/deceleration caused by wind and road resistance forces during inertial movement within a period $(t_i)$. The observations resulted in the equations below:

\begin{equation}
E_{b i}=-\delta \cdot G_a \cdot S_i \cdot\left(j_i-j_i^{P \Psi+P w}\right)(\mathrm{J}),
\label{eq3}
\end{equation}

where $\delta$ - rotational-inertia coefficient; $G_a$ - mass of the vehicle $(\mathrm{kg}), j_i$ - translational acceleration of the vehicle, m/s, $j_i^{P \Psi+P w}$-average acceleration/deceleration of the vehicle created by road and wind resistance forces during inertial movement (free rolling) in a period $\mathrm{t}_{\mathrm{i}}$ :

\begin{equation}
j_i^{P \Psi+P w}=-\left(b_0+b_1 \cdot V_1+b_2 \cdot V_1^2\right)\left(\mathrm{m} / \mathrm{s}^2\right),
\label{eq4}
\end{equation}

where $b_0$, $b_1$ and $b_2$ – equation coefficients which are calculated for each type of vehicle considering empirical road resistance force
coefficient $\Psi$.(Table \ref{tab1}).

\begin{table}[!h]
\caption{Values of indices for calculating $j^{P \Psi^{\prime}+P w}{ }_i$ of a vehicle with an average ($\Psi=0.015$)}
\begin{center}
\begin{tabular}{cccc}
\hline ${b}_0$ & ${~b}_1$ & ${~b}_2$ & $\delta$ \\
\hline 0.175 & $-5.63 \cdot 10^{-4}$ & $2.68 \cdot 10^{-4}$ & 1.05 \\
\hline
\end{tabular}
\label{tab1}
\end{center}
\end{table}

\subsection{Vehicle Characteristics}
In this study, 0, 50\% and 100\% CAV Market Penetration Rates (MPRs) have been tested. The intelligent driver model \citep{treiber2000congested} is used to model car-following behaviour. 
CAVs have reduced spacing and reaction time, half that of Human-Driven Vehicles (HDVs). Unlike CAVs, HDVs and AVs follow pre-trip shortest paths. HDVs calculate the shortest route at network entry and initial intersection based on prevailing traffic, with no further updates. HDVs are typical human-driven vehicles lacking communication capabilities, while AVs share CAV driving parameters but lack communication with traffic signals.
A combination of light passenger vehicles and heavy vehicles is incorporated into the traffic simulation.  

\subsection{Traffic Flow Characteristics}
In the study, the traffic flow outputs include several key factors based on the findings of previous studies \citep{alfaseeh2018impact}. The mean travel time and mean distance travelled, which convey that an increased MPR of CAVs led to reduced travel time and increased travel distances, particularly in congested conditions. The OD and flow patterns also showed that throughput improved with higher MPRs of CAVs, indicating quicker loading and unloading of traffic networks. Furthermore, average speed is calculated as an essential performance metric used for different steps of the simulation in the calculation of eco-driving, average travel time prediction, vehicle consumption model and emission prediction model. Lastly, the average density is another parameter calculated, which reveals less congestion with lower density over time, demonstrating the effectiveness of CAVs in facilitating quicker network unloading, particularly during peak hours.

\subsection{Emission model}
For the purpose of this study, the MOVES model \citep{liu2015more}, a creation of the United States Environmental Protection Agency, is utilized to gauge GHG emissions (expressed in CO2eq) along with NOx emissions. Mainly, GHG gases include Carbon Dioxide (CO2), Methane (CH4 ), and Nitrous Oxide (N2O).
According to \citet{althor2016global}, GHG gases, if combined together, are more harmful to the environment than just CO2. Hence, GHG in total was considered in this study instead of CO2 alone. CH4 and N2O account for climate change and have more adverse impacts than that of CO2. Therefore, their combined effect is converted and reported in terms of ‘‘CO2 equivalent’’ for GHG emission. The same approach is adopted for NOx, which has harmful effect on human body health \citep{United}. This justifies why CO2eq ER was selected as the modelling variable in our prediction models. To closely mirror local situations, input data are factored in, encompassing meteorology, fleet attributes, fuel type, and driving scenarios. Defining driving conditions entailed crafting operating mode (opMode) distributions, which denote the time a vehicle allocates to various operating modes. The opMode distribution offers valuable insights into vehicular behaviour within a road network, with the operating mode at any given second being dependent on the VSP. To each unique opMode, a specific emission rate is assigned, taking into account other elements like the model year of the vehicle and meteorology. Finally, total GHG and NOx emissions for each road segment are calculated using second-by-second traffic data for the 140 scenarios considered.
{The MOtor Vehicle Emission Simulator (MOVES) model calculates emissions based on the instantaneous speed and acceleration of vehicles} \citep{djavadian2020multi}. {MOVES uses second-by-second operating data, including speed and acceleration, to assign an appropriate power demand (specific power in kW/ton) and translate it into energy consumption in kWh for each type of vehicle.
Each vehicle in the simulation is assigned characteristics (such as vehicle type, model year, and weight and type of fuel), and the operating mode for each second is determined using the vehicle’s speed and acceleration. Based on the opmode, MOVES calculates the energy consumption for that specific time interval.
Once the energy consumption (kWh) is calculated for each vehicle based on their characteristics, the MOVES model estimates emissions such as CO2 and NOx. The emissions are computed in terms of grams of pollutants (e.g., CO2-equivalent, NOx) per vehicle and aggregated for the whole network. }

The simulation concludes upon all vehicles reaching their destinations. Space mean speed, density, and flow data for links are recorded every minute, an optimal update interval \citep{djavadian2020multi}.

For emission modelling, we adopt MOVES from the USEPA. It calculates GHG emissions (in CO2eq g/sec) and NOx using vehicle operating modes determined by VSP, as presented in Equation \ref{eq:10} for fossil fuel vehicles from \citet{alfaseeh2020greenhouse} and Equation \ref{eq:20} for electric vehicles from \citet{tu2020electric}.

\begin{equation}
    P_v,_t=\frac{A_1v_t+B_1v_t^2+C_1v_t^3+mv_ta_t}{m}  \label{eq:10}  
\end{equation}

\begin{equation}
    P_v,_t=\frac{A_2v_t+\frac{B_2}{v_t}+C_2v_t^2+mv_ta_t}{m}  \label{eq:20}  
\end{equation}

where:

\begin{math}
P_v,_t \end{math} is the vehicle specific power (VSP) at time t

\begin{math}
v_t \end{math} is the speed of vehicle at time t (m/sec)

\begin{math}
a_t \end{math} is the acceleration of vehicle at time t (m/sec$^2$)

m is the mass of vehicle, usually referred as ``weight'' (mg)

The coefficients A, B, and C pertain to constants linked to track-road, rolling resistance, rotational resistance, aerodynamic drag force, and vehicle tractive force.

Using vehicle emissions on a second-by-second basis for each link, the space mean GHG ER (measured in 
\begin{math}CO_2eq\end{math} g/sec) and NOx (in g/sec) for each link are computed through a predefined update interval.
It should be noted that energy consumption model for e-fuel vehicles are considered the same as ICEVs. Also, tailpipe emission from BEVs is deemed to be zero.

A prediction model for GHG, NOx and speed is required for future emission rate calculation for anticipatory eco-routing.
In previous studies, Long Short-Term Memory (LSTM) networks are applied to predict the amount of GHG, and average speed, in regard to taking the proactive single and multi-objective routing strategies into account \citep{alfaseeh2020deep}. We propose a Transformer network based prediction model, which outperforms the LSTM model proposed previously. 
%Our proposed transformer prediction model is retrained on high resolution datasets from real road network in the future by \citet{alfaseeh2020deep}. In this process, we applied the transformer neural network model, considering multi-objective routing strategies proactively. Compared to LSTM networks, the transformer model yielded superior results. This improvement is particularly noticeable in the predictions of speed and NOx emissions and to some extent in GHG emission predictions, where the Root Mean Square Error (RMSE), and correlation showed a notable improvement. The superiority of the transformer model can be attributed to its ability to handle long-term dependencies in time-series data more efficiently than LSTMs. Transformers leverage self-attention mechanisms, which weigh all other points in the sequence with respect to the current one, providing them with a global view of the data. This global context helps Transformers generate better predictions, particularly in complex scenarios where dependencies span long sequences.
{Instead of using the full Transformer for sequence-to-sequence tasks, we employ only the encoder part of the architecture, as our goal is to predict a single future value based on a historical sequence.
The model takes an input sequence of 250 time steps, with each time step representing a single scalar value (e.g., GHG emission value). This input is passed through a linear embedding layer to transform it from a scalar representation into a higher-dimensional space (250 dimensions). This transformation allows the model to work with the input more effectively within the Transformer framework. We used a fixed prediction window of 1 step. Thus, the model predicts the next time step (t+1) based on the previous 250 steps.
Since the Transformer does not inherently understand the order of time steps, we apply positional encoding to each input sequence. This encoding injects information about the position of each time step in the sequence by using sine and cosine functions at varying frequencies, which is standard in Transformer-based models.
The positional encoding ensures that the model has a sense of where each input time step lies relative to the others in the sequence, enabling it to make meaningful temporal predictions. We employ a Transformer encoder with multi-head self-attention to capture dependencies between time steps in the input sequence. This mechanism allows the model to attend to different parts of the sequence at each time step, which helps it identify patterns and relationships over different time scales.
The encoder uses 10 attention heads, which enables it to focus on multiple aspects of the time-series data simultaneously. The model consists of one encoder layer, which includes both multi-head self-attention and a feed-forward network. This architecture was chosen to balance model complexity with the size of the dataset.
After the input sequence has been processed by the Transformer encoder, the final output is passed through a linear decoder to map it back to a scalar value, which represents the predicted GHG value for the next time step.
The model is trained to minimize Mean Squared Error (MSE) between the predicted and actual values, as this is a standard loss function for regression tasks. The dataset is split into training and testing sets, with 80\% of the data used for training. The model is optimized using Adam optimizer, which is well-suited for this type of sequential data, and a learning rate scheduler is employed to decay the learning rate over time, improving convergence during training. Additionally, normalization techniques were applied to the input data to stabilize training and ensure better generalization.}

\begin{figure}
\centering
 \begin{subfigure}{0.64\textwidth}
     \centering
    \includegraphics[width=\textwidth]{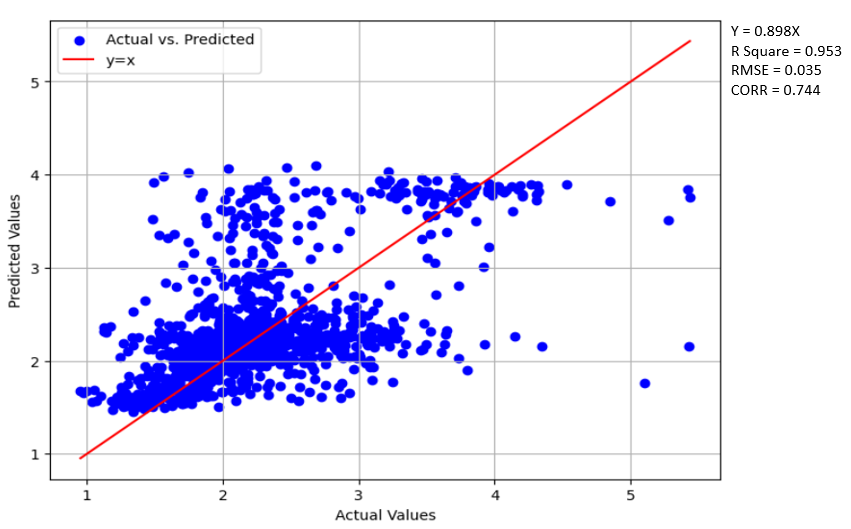}
     \caption{GHG ER(g/sec)}
     \label{fig:a}
 \end{subfigure}
% \hfill
 \begin{subfigure}{0.64\textwidth}
    \centering
     \includegraphics[width=\textwidth]{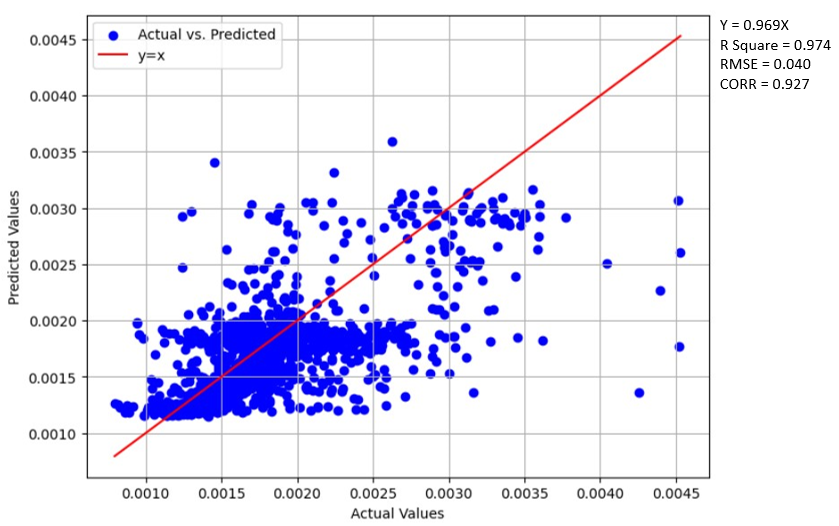}
     \caption{NOx ER(g/sec)}
     \label{fig:b}
 \end{subfigure}
 
% \medskip
  \begin{subfigure}{0.6\textwidth}
     \includegraphics[width=\textwidth]{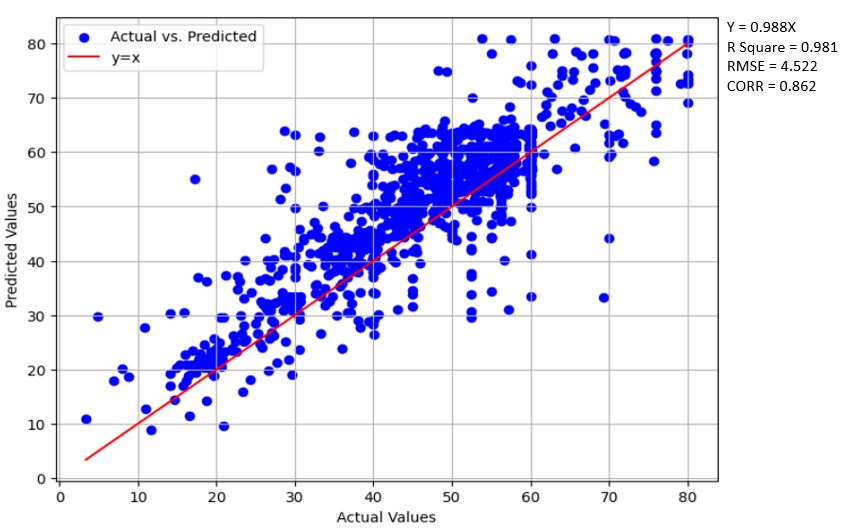}
     \caption{Speed (km/h) }
     \label{fig:c}
 \end{subfigure}
 \caption{Actual vs. predicted of the transformer models}
 \label{Label}
\end{figure}

Figure \ref{Label} presents our results, comparing true speed, NOx, and GHG emission rates from real data with the predictions generated by the Transformer model. The trend lines, nearly at 45-degree angles across all scenarios, indicate a good fit for the Transformer model, demonstrating its effectiveness in predicting emissions.
The Transformer model's superior performance in speed prediction compared to emission rate prediction can be attributed to the complexities of the quasi-convex relationship between GHG emission rates and their major predictors, such as speed, when compared to the relationship of speed predictor parameters, such as density. The Transformer model's ability to handle long-term dependencies in time-series data and capture intricate patterns makes it well-suited for accurate speed prediction, while the more intricate nature of emission rate prediction demands further investigation and fine-tuning.

The robustness of the Transformer model's predictions, as indicated by consistently low Root Mean Square Error (RMSE) values across different scenarios, underscores its reliability in capturing emission rate variations accurately. Its effectiveness in handling real-world data across various scenarios contributes to its suitability for prediction models in the transportation sector. The choice to use Transformer neural network modeling in our study was motivated by its proven success in natural language processing tasks and its capacity to capture long-range dependencies effectively. In the context of transportation data, which often involves time-series information and complex relationships between variables, the Transformer model's ability to consider distant interactions and dependencies is particularly valuable.
The Transformer model's architecture, with attention mechanisms, allows it to focus on relevant inputs, providing more contextually relevant predictions and accounting for intricate interactions between variables. This feature proved beneficial in our study, where emission rates are influenced by various factors that may be temporally distant from each other.
Additionally, the Transformer model's scalability and parallelization capabilities facilitate faster and more efficient computations, making it feasible for large-scale simulation-based analyses like ours. Its robustness and ability to handle diverse scenarios add to its appeal as a powerful tool for conducting comprehensive studies in the traffic prediction models domain.

\subsection{Upstream Cost of GHG and NOx}
The Greenhouse Gases, Regulated Emissions, and Energy Use in Transportation (GREET) model \citep{dai2019life}, developed by the U.S. Department of Energy, is utilized to calculate well-to-wheel (WTW) GHG and NOx emissions for various fuel types. The main results exclude vehicle cycle emissions (material production, manufacturing, assembly, maintenance, disposal) due to a lack of Canada-specific data. Default values in the GREET model compute impacts of the vehicle cycle and other inputs. Unutilized life cycle indicators like water consumption and land use are omitted, given their irrelevance to Canadian and U.S. EV development policy-making and lack of available data.

NOx emissions, primarily from power generation and industrial boilers, are updated in the current study based on 2022 unit-specific data \citep{kelly2022cradle}. BEVs are assumed emission-free at the operational level. ICEVs and HEVs operational level emissions are determined based on simulator prediction elaborated in previous sections.

We have also used the fuel consumption values for BEV from \citet{kelly2022cradle} in which long-range BEVs are chosen to 
decrease the range discrepancy with other vehicle fuel technologies. 
Other inputs of upstream emission from e-fuels, BEVs, HEVs, and ICEVs are also taken from \citet{kelly2022cradle}, \citet{rossi2022analysis} and \citet{elgowainy2016cradle}.

\subsection{Generalized Cost for Multi-Objective Optimization}
Equation \ref{eq:3} defines the multiple objectives in the optimization process. All the costs associated with the objectives are converted into equivalent Canadian dollar values and added to the objective function.

\begin{equation}
\begin{split}
  min  \sum_{l}^{} \lbrace \hat{T}_{(l,\triangle_j)} \times  \beta_T \times W_T &+ E_{\hat{C}O_{2(l,\triangle_j)}} \times \beta_{CO_2} \times W_{CO_2} \\ 
  &+ E_{\hat{N}O_{x(l,\triangle_j)}} \times \beta_{NO_x} \times W_{NO_x} \rbrace
  \label{eq:3}  
\end{split}
\end{equation}

The approximate GHG costing method, which is developed and investigated by \citet{alfaseeh2020deep}, is used to measure the amount of GHG cost. Equation \ref{eq:4} and \ref{eq:5} represent the GHG and NOx cost of each link, which is related to the link Travel Time (TT). Equation \ref{eq:6} also estimates the TT, where $D_l$ represents link $l$ length and $v_l,_{\Delta_j}$ is link $l$ speed of $_{\Delta_j}$  time interval.

\begin{equation}
    GHG cost_{(l,{\Delta_j})}=GHG_{ER(l,\Delta _j)} \times  TT_{(l,\Delta_j)} \label{eq:4}  
\end{equation}

 \begin{equation}
    NO\text{x} cost_{(l,\Delta_j)}=NOx_{(l,\Delta_j)} \times  TT_{(l,\Delta_j)} \label{eq:5}  
\end{equation}

\begin{equation}
    TT_{(l,\Delta_j)}=\frac{ D_l}{ v_{(l,\Delta_j)}}  \label{eq:6}  
\end{equation} 
 
Equations \ref{eq:4}, \ref{eq:5}, and \ref{eq:6} are used to calculate GHG and NOx, by using the weighted average of emissions to provide higher weight to the latest seconds of intervals. At the end of this section, the total amount of GHG and NOx emission cost over the entire network can be estimated. Therefore, an available network level emission cost allows the fuel-based scenarios to be compared. It should be noted that multi-objective routing calculates the cost of optimized objectives in Equation \ref{eq:3} and normalize the costs based on the defined objective \citep{alfaseeh2019multi}.

\subsection{Other Costs}
The comparison among several fuel combinations in our network takes into account various costs other than just NOx and GHG emissions, including travel time cost, fuel cost, average operation and maintenance cost, and average vehicle price.
\begin{itemize}
    \item \emph{Travel Time Cost:} This is a significant expense in transportation, and it varies depending on various factors such as vehicle type, time of day, and traffic conditions. Toronto's average travel time cost is considered 27.66 CAD in the year 2021 by the Statistics Canada report \citep{CanadaStatistics} {as it was estimated to be about 55 minutes of time and the average Canadian salary in 2021 was about \$30 per hour}.   
    
    \item \emph{Fuel Cost:} In our analysis conducted in Ontario, we calculated the average fuel costs for sedan and SUV models across various fuel options: internal combustion engine vehicles, battery electric vehicles, hybrid electric vehicles, and e-fuels based on resources from \citet{stringer2022assessing}. SUVs generally have higher fuel consumption and costs compared to sedans due to their larger size in all fuel types. In contrast, BEVs rely solely on electricity and offer lower day-to-day fuel expenses compared to ICEVs. HEVs provide better fuel efficiency through hybrid technology, while e-fuels, synthetic fuels from renewable sources, offer reduced carbon emissions. Using the distribution of sedan and SUV models in Ontario, we determined the average fuel costs for each vehicle type and fuel option in our analysis.
    It's  also important to consider factors such as variation in fuel prices, vehicle efficiency, driving conditions, and charging/refuelling infrastructure for alternative options like BEVs and e-fuels which are not the scope of this study. 

    \item \emph{Operation and Maintenance Cost:} In this study the operation and maintenance cost is also calculated with the same method as the fuel price, elaborated in the previous section. These costs can vary significantly depending on the type of vehicle. ICEVs usually have higher maintenance costs due to their complex mechanical systems. BEVs, in contrast, have fewer moving parts and therefore usually incur lower maintenance costs. HEVs may fall somewhere in between due to their dual powertrain. E-fuels can also influence the operation and maintenance costs depending on their compatibility with existing engines \citep{stringer2022assessing}.

    \item \emph{Vehicle Customer Price:} The average price of passenger vehicles varies depending on the type of vehicle (e.g., SUV, sedan, etc.) and the propulsion technology (BEV, HEV, ICEV, etc.). BEVs and HEVs typically have a higher upfront cost than ICEVs, but can potentially offset this with lower operating costs. The vehicle customer price is also considered in the cost analysis of this study \citep{axsen2022comparing}. For zero-emissions vehicles, we are assuming that the price includes the rebates offered by the Federal and Provincial governments.
    
\end{itemize}

\begin{landscape}
\begin{figure}
\centering
 \begin{subfigure}{0.5\textwidth}
     %\centering
    \includegraphics[scale=0.65]{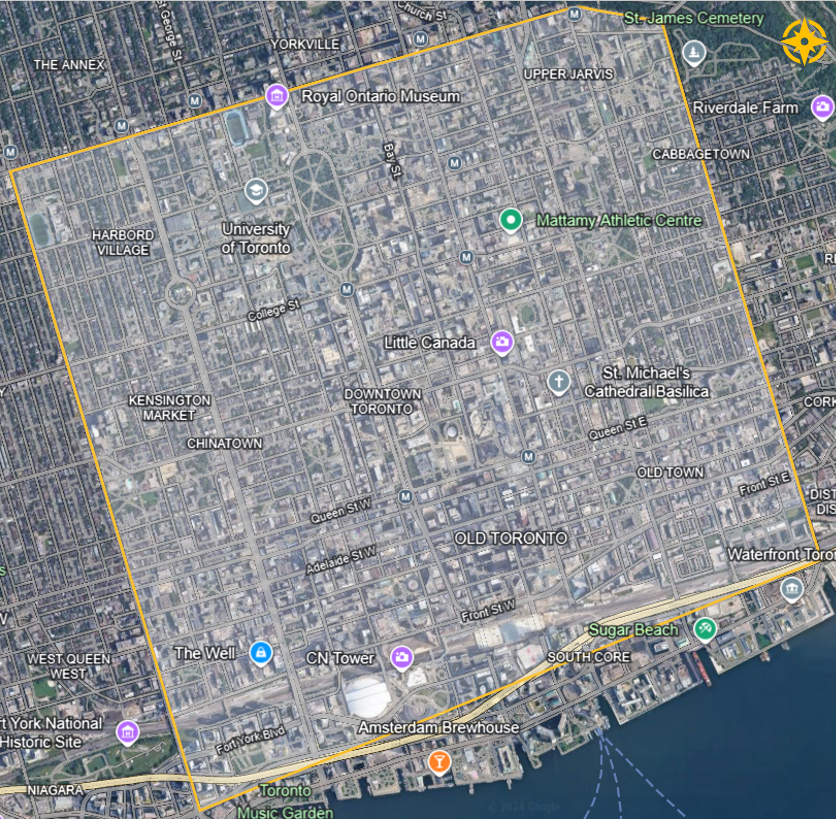}
   \caption{{Actual (Scale 0.001)}}\label{dt1}
 \end{subfigure}
% \hfill
 \begin{subfigure}{0.45\textwidth}
    %\centering
    \hspace{1.75cm}
      \includegraphics[scale=0.83]{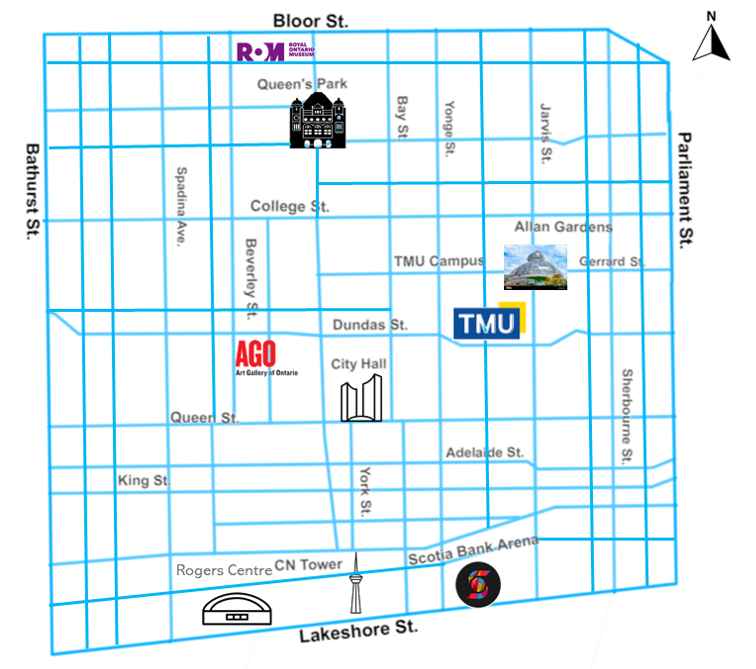}
   \caption{Simplified and tilted}\label{dt2}
 \end{subfigure} 
 \caption{Downtown Toronto Street Network-Actual vs. Simplified}\label{dt}
\end{figure}
\end{landscape}

%In this segment, a brief overview of the selected study area and the generated data is provided. We also outline the parameters configured for the simulation within the downtown Toronto network. 
The study site selected for this study is downtown Toronto, which is chosen due to its recurrent peak hour congestion in both mornings and afternoons. Figure \ref{dt} shows a network of 268 nodes (intersections) and 839 links (road segments) within a 3.14 km x 3.31 km area in the downtown core. The study employed an agent-based microsimulator for traffic and emission analysis, generating high-resolution GHG and traffic flow data for links \citep{FarooqBilalandDjavadian}. Vehicular movement used a calibrated Intelligent Driver Model (IDM) \citep{treiber2000congested}. Vehicles operated from 7:45 to 8:00 A.M., dynamically navigating network links using real-time traffic data until destination. City of Toronto's Transportation Tomorrow Survey (TTS) travel data with a 2021 post-pandemic growth factor simulated demand.

\section{Results and Analysis}
\label{res}
This section presents the findings from the comprehensive simulation scenarios explained in the previous section, focusing on the comparison of GHG emissions, NOx emissions, travel time, and all relevant costs across different scenarios with varying fuel type combinations and CAV penetration rates. %In the results section, we provide a detailed analysis of the scenarios, each representing a unique combination of fuel types, routing strategy, and CAV integration levels. 
The description of the convention used to label the scenarios is as follows:

\begin{itemize}
    \item 0\% CAV: The routing strategy investigated in this scenario is User Equilibrium (UE), where all vehicles, including conventional vehicles, follow the UE approach for route selection.
    \item 50\% CAV: In this scenario, it is assumed that 50\% of the vehicles, which are conventional vehicles, adopt the UE routing strategy, while the other 50\% of vehicles, which are CAVs, utilize either the Myopic (M) or Anticipatory (A) routing strategy.
    \item 100\% CAV: In this scenario, it is assumed that all vehicles, both conventional and CAVs, follow either the Myopic (M) or Anticipatory (A) routing strategy.
\end{itemize}

For all scenarios, two different driving conditions are considered: one where drivers practice eco-driving (ED) techniques and another where eco-driving is not applicable (NED).
Additionally, in the mixed-fuel type condition, we examine 14 different combinations of mixed-fuel environments across the scenarios. These combinations encompass various fuel types and their respective proportions within the vehicle fleet as displayed in Table \ref{tab2}. The detailed analysis of each scenario provides insights into the impacts of fuel type combinations, CAV penetration rates, and eco-driving practices on GHG emissions, NOx emissions, and travel time. By examining these scenarios, we aim to contribute to the understanding of the most effective pathways for reducing emissions and optimizing travel time in mixed-fuel environments while considering various costs. 

%The findings from this study have implications for transportation planning, policy-making, and the advancement of sustainable and efficient transportation systems.
\begin{table}[h]
\caption{Fuel Combination Scenarios Description} 
\centering
\begin{tabular}{|l|c|c|c|c|}
\hline & ICEV\% & HEV\% & BEV\% & e-fuel\% \\
\hline I100 & 100 & 0 & 0 & 0 \\
\hline I50H50 & 50 & 50 & 0 & 0 \\
\hline ES & 25 & 25 & 25 & 25 \\
\hline $125 \mathrm{H} 75$ & 25 & 75 & 0 & 0 \\
\hline $125 \mathrm{~B} 75$ & 25 & 0 & 75 & 0 \\
\hline $175 \mathrm{H} 25$ & 75 & 25 & 0 & 0 \\
\hline $175 \mathrm{~B} 25$ & 75 & 0 & 25 & 0 \\
\hline B100 & 0 & 0 & 100 & 0 \\
\hline E100 & 0 & 0 & 0 & 100 \\
\hline I50B50 & 50 & 0 & 50 & 0 \\
\hline I50E50 & 50 & 0 & 0 & 50 \\
\hline I25E75 & 25 & 0 & 0 & 75 \\
\hline H100 & 0 & 100 & 0 & 0 \\
\hline I75E25 & 75 & 0 & 0 & 25 \\
\hline
\end{tabular}
\label{tab2}
\end{table}

\subsection{GHG Emission}
In Figure \ref{Labelghg}, GHG emission based on CO$_2$ eq kg is displayed for all CAV MPRs, including 0\%, 50\%, and 100\%. In Figure \ref{fig:d}, eco-driving has decreased the GHG emission by about 5-10\% in all the UE scenarios. 
For all three CAV penetration rates, the 100\% BEVs has the least WTW GHG pollution, mainly because it does not have any tailpipe emissions \citep{logan2021phasing}. 
In the second and third place, there come the scenarios with 75\% BEVs, and 50\% BEV. However, the BEVs are considered a costly alternative, with limited production and long delivery wait times, which makes them an unaffordable option, at least in the foreseeable future, for most consumers. Therefore, looking at other green alternatives is important in planning for more sustainable transportation. 100\% HEV decreased GHG emissions by about 40\%, if combined with eco-driving, and by 35\% with no eco-driving. With 50\% HEVs in the network, the decrease in GHG emission is about 15\%. The other green option, e-fuels, has the potential to reduce the emission in the network by about 5\%, which is not considerable for passenger vehicles. An equal share of all four fuel types can reduce the WTW GHG emission by about 25\%, which can be a realistic goal that accounts for the preference heterogeneity among the population. 

In Figure \ref{fig:e}, all mixed-fuel scenarios are investigated with/without eco-driving and with 50\% of vehicles, as CAVs, applying Myopic (M) or Anticipatory (A) eco-routing, while the other half, as human driven vehicles, routing based on User Equilibrium (UE). Overall, it can be understood that anticipatory routing has the potential to reduce GHG emission by about 8-10\% compared with myopic routing. 50\% CAVs with anticipatory eco-routing, is capable of mitigating GHG emission by about 10-12\% over all scenarios compared with UE. It can also be seen that considering anticipatory eco-routing for 50\% CAVs, 75\% BEV in the network is greener than 100\% BEV regardless of eco-driving. It is mainly due to the high amount of upstream emission of BEVs.  The rest of GHG emission ranking stays the same as non-CAV environment.

In Figure \ref{fig:f}, 100\% of vehicles are CAV and the routing strategy is also either myopic or anticipatory. The same as 50\% CAVs, anticipatory routing can reduce the GHG emission by about 8-10\% more than myopic. Also, anticipatory routing combined with eco-driving has the ability to decrease GHG emission by about 15-17\% compared with UE strategy with eco-driving and 25-27\% compared with UE strategy with no eco-driving. It should also be noted that in Figure \ref{fig:f} similar to Figure \ref{fig:e}, with all vehicles using anticipatory routing, 75\% BEV is greener than 100\% BEV if WTW emission is considered.
It is worth mentioning that in all cases in Figure \ref{Labelghg},  the total GHG emission of 100\% ICEVs scenario, in which the network only contains fossil fuel vehicles is the highest. This can clarify the significance of moving towards alternative fuels in transportation. 
The specific values observed in the GHG emission results can be attributed to the unique characteristics of each scenario. For instance, in the 100\% CAV scenario with BEVs, the absence of tailpipe emissions leads to the least WTW GHG pollution, which results in the lowest GHG emission value. On the other hand, the 100\% ICEVs scenario, with only fossil fuel vehicles, has the highest total GHG emissions due to the lack of green alternatives. The varying reduction percentages in GHG emissions under different scenarios reflect the effectiveness of eco-driving and anticipatory routing in optimizing vehicle behaviour and reducing emissions. Furthermore, the ranking changes observed in certain scenarios highlight the significance of anticipating eco-routing for certain vehicle combinations, like the 75\% BEV scenario outperforming the 100\% BEV scenario in terms of WTW emissions when utilizing anticipatory routing.

In terms of comparative analysis among the scenarios (a, b, and c), we can observe how the different strategies impact GHG emissions. For example, comparing the GHG emission reductions achieved through eco-driving and anticipatory routing, we see that the latter consistently outperforms myopic routing in all scenarios. This demonstrates the importance of considering advanced routing strategies when trying to optimize emissions. Additionally, the percentage reductions in GHG emissions can be compared across different CAV penetration rates. This allows us to gauge how much impact CAVs have on emission reductions, and how this impact scales with increasing CAV penetration.
 These results provide valuable insights for different stakeholders. Policymakers can use them to identify effective strategies for reducing emissions, such as promoting the adoption of CAVs with anticipatory eco-routing and implementing eco-driving practices. Urban planners can leverage these findings to design more sustainable transportation networks with a mix of green vehicles. Manufacturers and technology developers can assess the potential market demand for different fuel technologies and focus on cost-effective solutions. Additionally, these results are relevant for consumers, as they highlight the environmental benefits and cost implications associated with different vehicle choices, influencing their preferences towards more eco-friendly options.
 
{The results demonstrate that BEVs, particularly when combined with eco-driving and anticipatory routing strategies, achieve the highest reduction in greenhouse gas (GHG) emissions, with a decrease of 25-27\% in Well-to-Wheel (WTW) emissions. This aligns closely with global studies, such as the one by the International Council on Clean Transportation (ICCT), which found that BEVs globally reduce lifecycle GHG emissions by 66-69\% compared to gasoline vehicles, and up to 78\% in regions with cleaner energy grids, such as the European Union \citep{bieker2021global}. Similarly, the European Environment Agency (EEA) reported an 18-25\% reduction in lifecycle emissions for BEVs compared to ICEVs. These reductions support the view that BEVs, while costly, offer substantial GHG benefits when integrated into advanced urban transport systems \citep{gryparis2020electricity}.
In another similar study for CAVs, With very high traffic demand in a smaller network, 100\% CAV with Anticipatory routing decreased total GHG emissions by around 40\% compared to the 0\% CAV scenario \citep{tu2019quantifying}. In the current study, in a larger scale network, the results show about 20-25\% improvement in GHG in similar scenarios.}

{\citet{kato2016ecodriving} showed that if Eco-driving is implemented under low demand (ICEVs and HEVs), CO2-reduction rates ranged from 10.9\% to 12.6\%, respectively. For BEVs, CO2-reduction rates were higher, ranging from 11.7\% up to 18.4\%.
Our study also indicates that EVs have a higher potential for CO2 reduction when maintaining high energy conversion efficiency by about 9-10\% in various combinations of ICEVs, HEVs, and BEVs.}

\begin{figure}
\vspace{-1cm}
\centering
 \begin{subfigure}{0.7\textwidth}
 \includegraphics[width=1.2\textwidth]{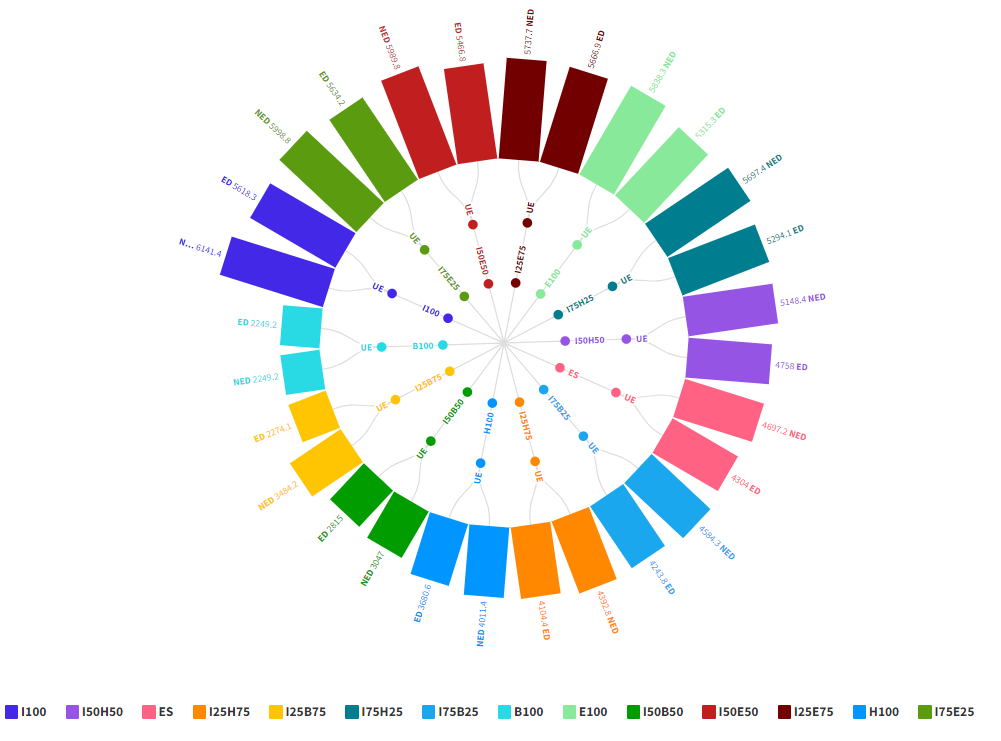}
     \caption{0\% CAV with UE routing}
     \label{fig:d}
 \end{subfigure}
 %\hfill
  \begin{subfigure}{0.7\textwidth}
     \includegraphics[width=1.2\textwidth]{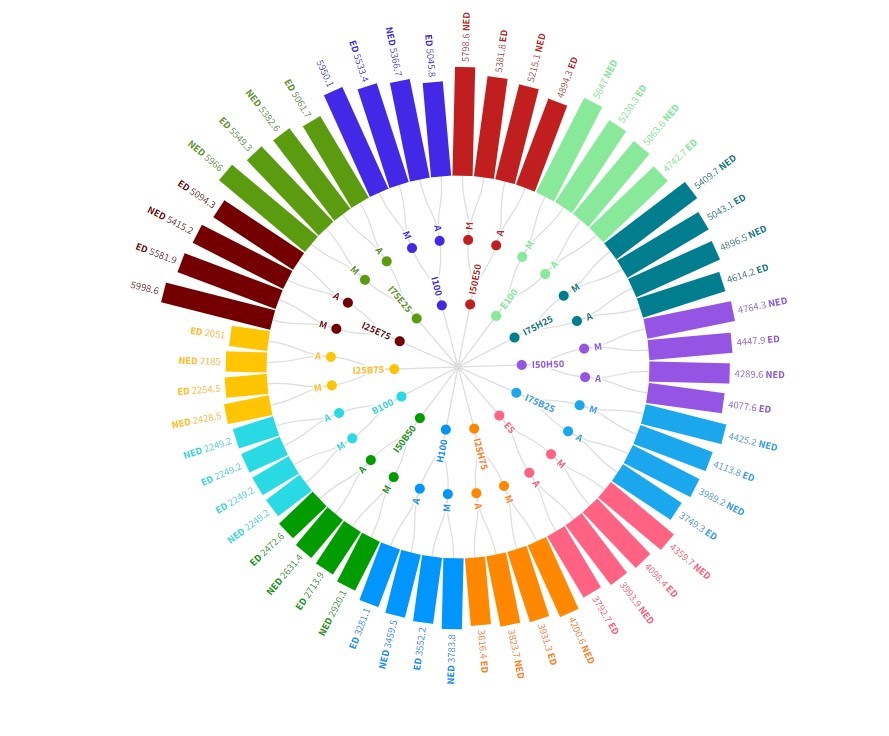}
     \caption{50\% CAV with eco-routing}
     \label{fig:e}
 \end{subfigure}
\end{figure}
% \pagebreak
%\newpage
%\medskip

\begin{figure}
\ContinuedFloat
    \centering
 \begin{subfigure}{0.7\textwidth}
     \includegraphics[width=1.2\textwidth]{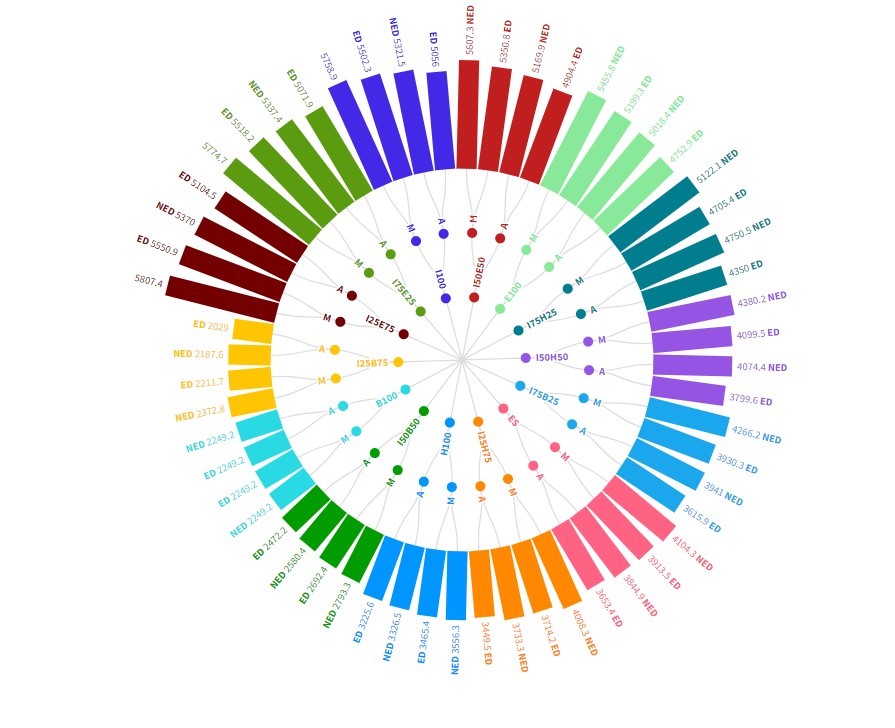}
     \caption{100\% CAV with eco-routing}
     \label{fig:f}
 \end{subfigure}

 \caption{GHG Emission (CO$_2$ eq kg)}
 \label{Labelghg}
\end{figure}

\subsection{NOx Emission}
The adoption of different vehicle technologies, such as battery electric vehicles (BEVs), hybrid electric vehicles (HEVs), and e-fuels, can have a significant impact on Nitrogen Oxides (NOx) emissions, both at the tailpipe and well-to-wheel (WTW) levels. Additionally, integrating anticipatory routing and eco-driving practices can further influence emissions in various scenarios.
In Figure \ref{Labelnox}, NOx emission based on CO$_2$ eq kg is displayed for three MPR of CAVs including 0\%, 50\%, and 100\%.  
In the context of NOx emissions, the specific values of different scenarios are influenced by the choice of vehicle technology and the adoption of eco-driving and routing strategies. 
In terms of NOx, BEVs have much less upstream emission than other alternatives, making them a prominent winner in the competition of greener fuel technologies. The inclusion of other fuel types (HEVs and e-fuels) in the network leads to higher NOx emissions, even with eco-driving and routing strategies. Considering NOx emission, eco-driving has shown to reduce the emission by about 3-5\% regardless of routing strategy (UE, M, and A). Furthermore, Anticipatory routing in CAVs has proven to decrease NOx emission by 5-7\% without eco-driving and 7-9\% with eco-driving.
The implementation of eco-driving practices is observed to contribute to a modest reduction in NOx emissions, regardless of the routing strategy used (UE, M, and A). However, the most significant NOx emission reduction is achieved when anticipatory routing is combined with eco-driving practices. Anticipatory routing optimizes CAVs' routes efficiently, leading to smoother traffic flow and less idling, which results in notable NOx emission reductions.
In all three CAV MPRs, 100\% BEV is considered the best choice, and 75\% BEV falls after that. However, it should be noted that the overall NOx emissions associated with BEVs depend on the electricity generation mix. In regions with a high proportion of renewable energy sources, the WTW NOx emissions can be significantly lower compared to regions that heavily rely on fossil fuels. Moreover, as said before, BEVs are a costly and out of reach option for most users, therefore, an equal share of all four fuel types can be a more realistic alternative, decreasing NOx emission by about 25\%. In addition, 100\% of HEVs in the network can reduce NOx by about 60\% which is a giant leap toward improving health quality indices of city dwellers.
In light of these findings, it becomes evident that BEVs have a significant advantage in terms of NOx emissions reduction, but practical constraints like cost and production capacity may limit their widespread adoption. Therefore, a balanced approach, such as an equal share of all four fuel types, emerges as a more realistic and effective strategy for achieving substantial NOx emission reductions while considering the challenges and preferences of different stakeholders. Additionally, the implementation of anticipatory routing and eco-driving practices further enhances NOx emission reductions across all scenarios, emphasizing their importance in shaping sustainable transportation strategies. {It should be highlighted that while EVs perform well in reducing CO2 emissions, their environmental impact is more complicated when considering other pollutants like NOx and N2O due to the upstream emissions from electricity production \citep{veza2023electric}.

\begin{figure}
\centering
 
 \begin{subfigure}{0.7\textwidth}
     \includegraphics[width=1.2\textwidth]{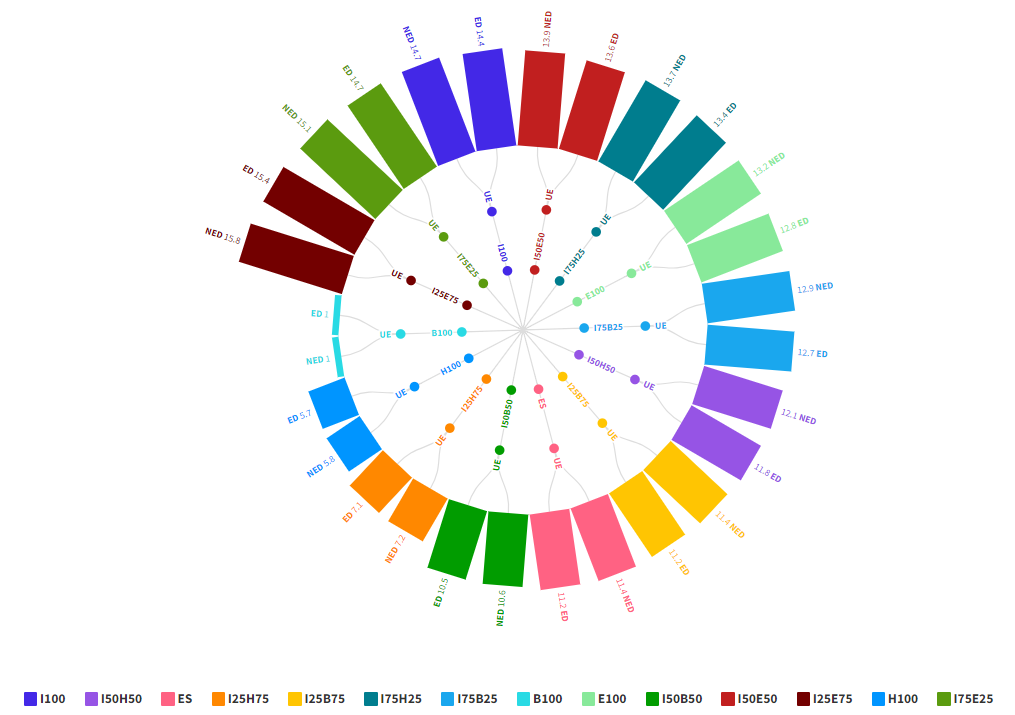}
     \caption{0\% CAV with UE routing}
     \label{fig:g}
 \end{subfigure}
% \hfill
  \begin{subfigure}{0.7\textwidth}
     \includegraphics[width=1.2\textwidth]{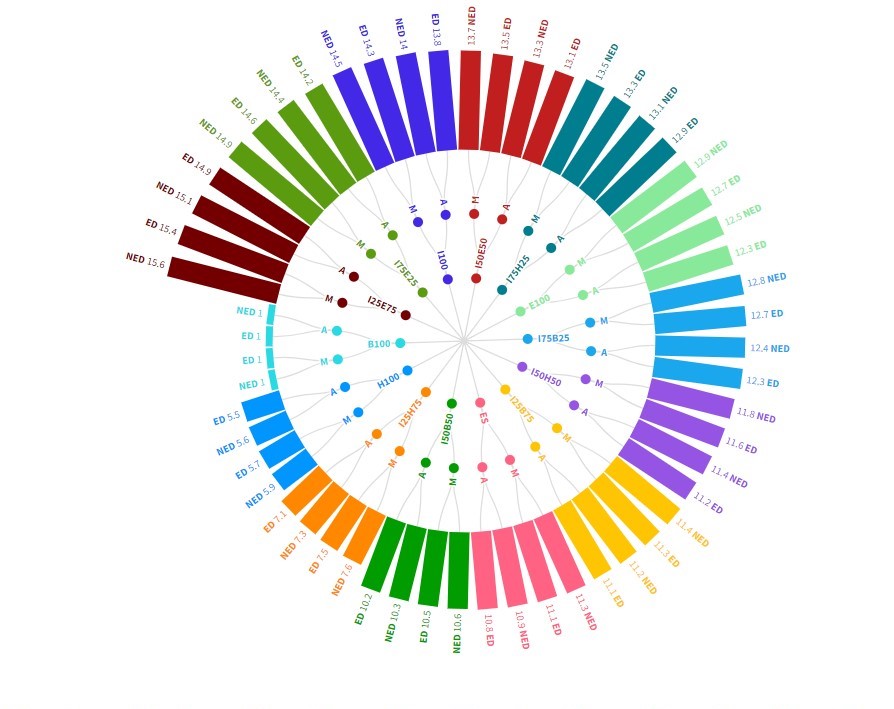}
     \caption{50\% CAV with eco-routing}
     \label{fig:h}
 \end{subfigure}
\end{figure}
% \pagebreak
%\newpage
%\medskip
\begin{figure}
\ContinuedFloat
    \centering
 \begin{subfigure}{0.7\textwidth}
     \includegraphics[width=1.2\textwidth]{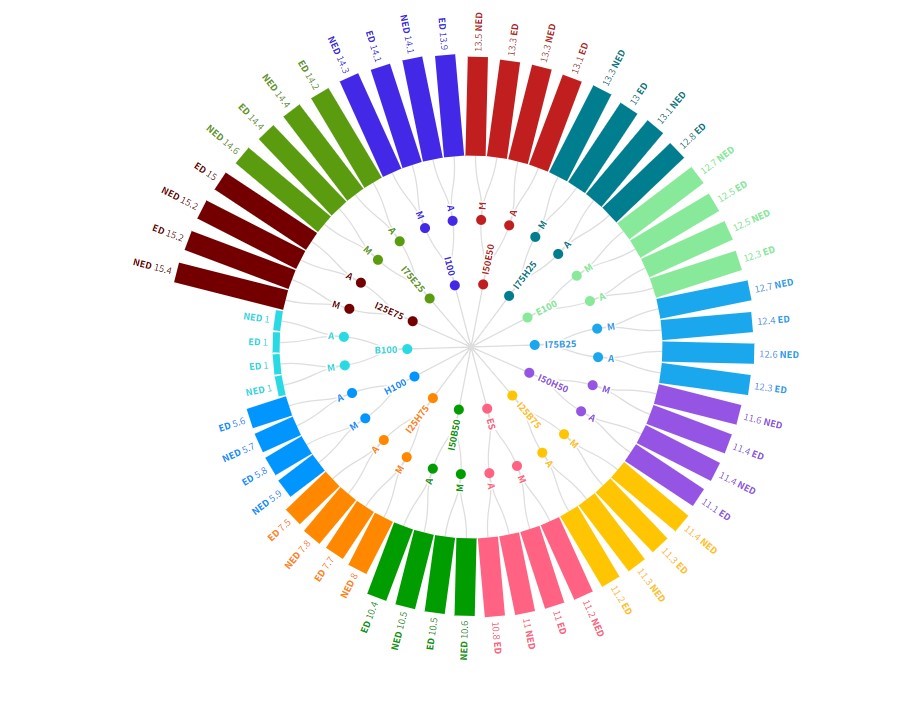}
     \caption{100\% CAV with eco-routing}
     \label{fig:i}
 \end{subfigure}

 \caption{NOx Emission (CO$_2$ eq kg)}
 \label{Labelnox}
\end{figure}

\subsection{Travel Time}
Average travel time in the network, can be influenced by various factors, including traffic congestion, road conditions, route selection, and driving behaviour. The integration of different vehicle technologies and strategies can have an impact on travel time. In terms of the impact of fuel type on average travel time, BEVs typically have a limited driving range compared to conventional vehicles. This may necessitate more frequent stops for recharging, potentially increasing travel time for longer journeys. However, advancements in charging infrastructure and longer-range BEVs are helping to mitigate this limitation. HEVs combine the benefits of both internal combustion engines and electric motors, allowing for improved fuel efficiency. This can result in reduced refuelling stops and potentially shorter travel times compared to conventional vehicles. Vehicles running on e-fuels generally exhibit similar performance characteristics to those running on conventional fossil fuels. Thus, travel time is comparable to that of traditional internal combustion engine vehicles. In our study, refuelling is not considered as the simulation is only 15-minute intervals during the morning peak hour of the downtown Toronto network. Therefore, the fuel type impact on travel time is negligible in our scenarios. 

The choice of routing strategy can have a significant impact on travel time. Myopic routing takes into account factors such as distance, road capacity, and congestion levels to minimize travel time. By selecting the shortest or fastest routes based on the current situation, myopic routing can potentially reduce travel time in the short term. By taking into account upcoming traffic conditions, anticipatory routing aims to avoid congested areas and optimize travel routes for fuel efficiency and reduced travel time. It can potentially provide better travel time performance by minimizing the impact of congestion and delays. In a user equilibrium scenario, each traveller makes routing decisions based on their own self-interest, aiming to minimize their individual travel time. The equilibrium reflects the collective result of individual route choices. 
In addition, eco-driving techniques, such as maintaining a steady speed and avoiding sudden accelerations or decelerations, can improve energy efficiency and reduce travel time by promoting smoother and more efficient driving.
According to Figure \ref{fig:j} eco-driving has the potential to decrease travel time by about 7-9\% while HDVs are taking UE routing strategy. Based on Figure \ref{fig:l}, 100\% CAVs with anticipatory routing, can reduce TT by about 20\% with no eco-driving, and by about 25\% if combined with eco-driving techniques. In addition, according to Figures \ref{fig:l},and \ref{fig:k}, travel time with anticipatory routing is about 7-9\% less than myopic routing.

{{A CAV-focused study by \citet{samaranayake2024impact} demonstrates that travel time decreases significantly as CAV market penetration rate increases. Specifically, the weighted average travel time decreases by 28\% as the MPR of CAVs reaches 100\%. \citet{djavadian2020multi} reported similar imporovements. In our study, CAVs with anticipatory routing also substantially reduced travel time, with a 20\% decrease in travel time without eco-driving techniques and 25\% reduction when eco-driving is applied. This suggests that anticipatory routing, which proactively considers future traffic conditions, complements CAVs' benefits by further optimizing routes and reducing travel times.}}

\begin{figure}
\vspace{-1cm}
\centering
 \begin{subfigure}{0.7\textwidth}
 \includegraphics[width=1.2\textwidth]{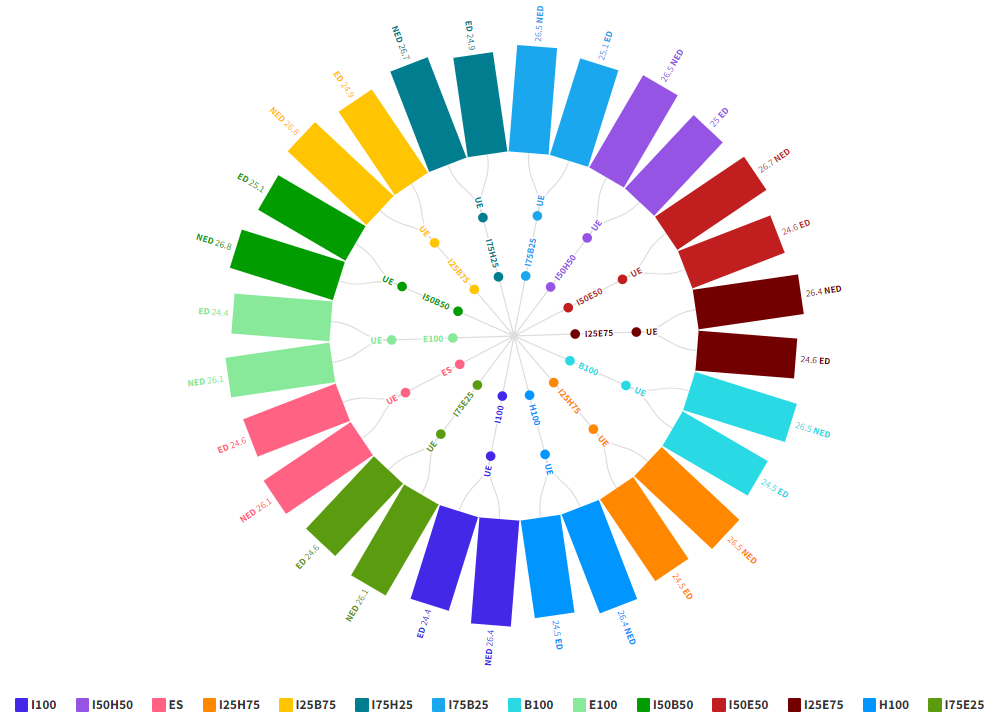}
     \caption{0\% CAV with UE routing}
     \label{fig:j}
 \end{subfigure}
 %\hfill
  \begin{subfigure}{0.7\textwidth}
     \includegraphics[width=1.2\textwidth]{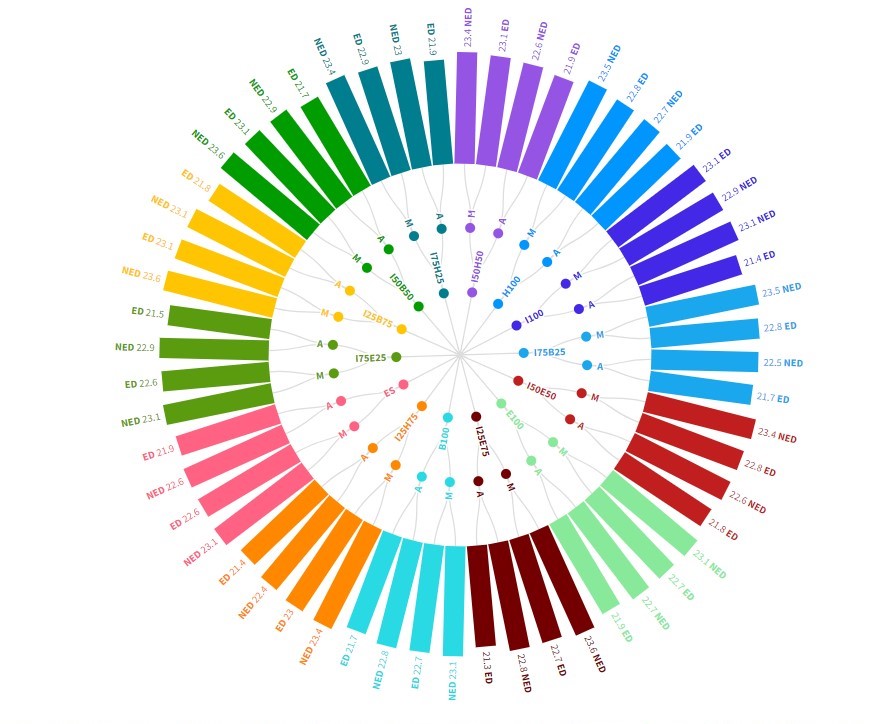}
     \caption{50\% CAV with eco-routing}
     \label{fig:k}
 \end{subfigure}
\end{figure}
% \pagebreak
%\newpage
%\medskip

\begin{figure}
\ContinuedFloat
    \centering
 \begin{subfigure}{0.7\textwidth}
     \includegraphics[width=1.2\textwidth]{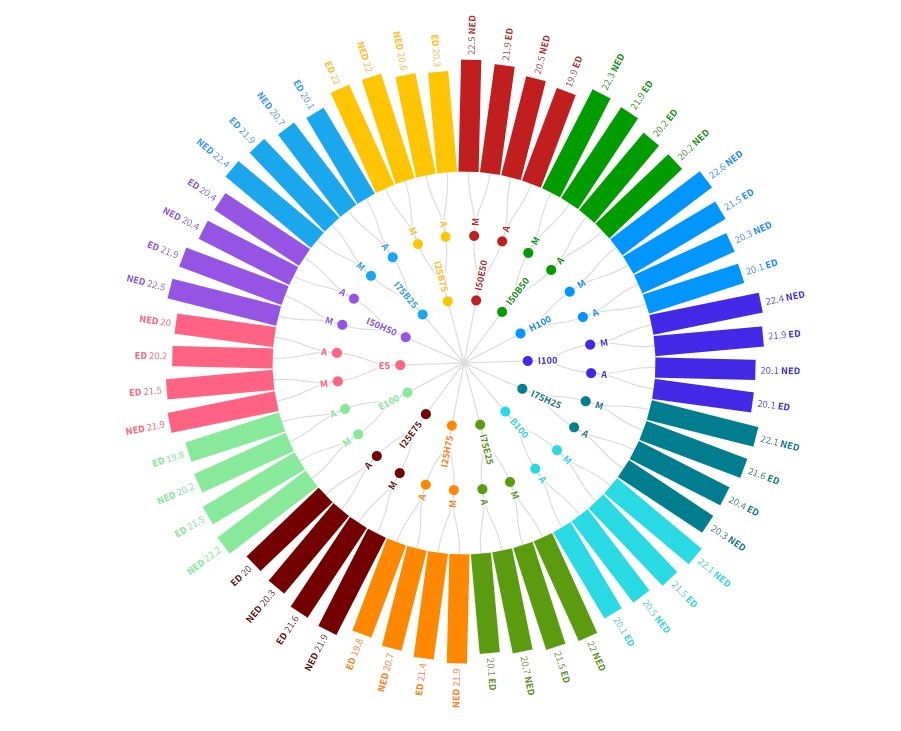}
     \caption{100\% CAV with eco-routing}
     \label{fig:l}
 \end{subfigure}

 \caption{Travel Time (Minutes)}
 \label{Labeltt}
\end{figure}

\subsection{Cost Analysis}
Conducting a lifecycle cost analysis based on NOx and GHG emission costs involves considering the financial implications associated with the emissions of nitrogen oxides (NOx) and greenhouse gases (GHGs) throughout the entire lifecycle of different technologies and scenarios based on the previous parts of the study. We have divided the costs in the unit of CAD per kilometer,  based on emission costs and other costs including travel time costs, fuel costs, vehicle prices and operation and maintenance costs.

{Figure \ref{fig:O} is the Pareto chart for the feature importance analysis of the regression model used for finding the importance of each parameter in the 140 scenarios analysed in this study. In this figure, the bars represent the individual importance of each feature and the line with diamond markers indicates the cumulative importance of the features in percentage terms. It demonstrates how different features contribute to the overall importance in predicting the Emission related cost, with a clear visualization of which features have the most significant impact. The figure highlights that a small number of features (particularly BEV and HEV) were responsible for the majority of the influence on the emission cost. The analysis provided valuable insights into the factors affecting emission costs. The dominance of vehicle type in the feature importance suggests that strategies focusing on the promotion and adoption of electric and hybrid vehicles could be effective in managing GHG costs.}
\begin{figure}
    \centering
     \includegraphics[width=\textwidth]{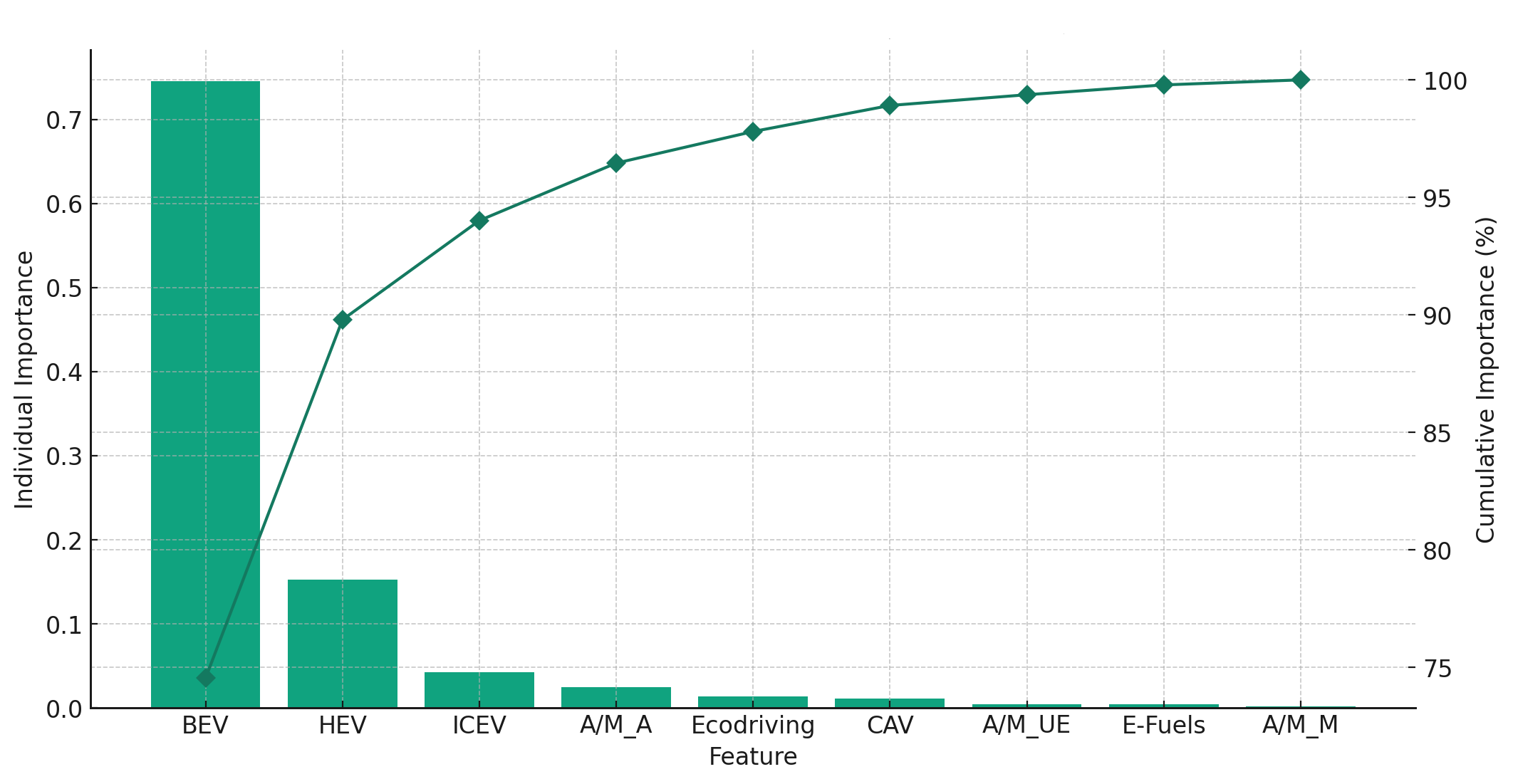}
     \caption{Emission-related costs Pareto analysis}
     \label{fig:O}
 \end{figure}

The cost analysis in Figure \ref{costs}, refers to the emission costs (E) and Other costs (NE) for the scenario of 100\% CAVs in the network. In the previous subsections of the results, we found that the combination of CAV MPRs, eco-driving, and routing strategy had negligible impact on the ranking of fuel combinations in terms of emissions. Therefore, we focused our cost analysis on the 100\% CAV MPR scenario, as the ranking remained unchanged for the 0\% and 50\% CAV scenarios when considering eco-driving and eco-routing. To visually represent the cost analysis, we reckoned that emission costs and non-emission costs both play decisive roles for customers. However, we observed that emission costs are much higher than non-emission costs. To address this discrepancy, we assigned equal weights to both factors. For each type of cost, we assigned a score of 50 to the best option (least expensive), and the other alternatives received scores based on their relative weight (cost) compared to the best option. Based on this approach, it is found that 100\% BEV received a score of 50 in emission-related costs, while 100\% ICEV received a score of 50 in other-related costs. When considering non-emission costs, 25\% HEVs and 25\% e-fuels ranked second and third, respectively. This highlights the need for significant cost reductions in BEVs to make them more competitive in terms of vehicle price, fuel, and maintenance costs (The score is out of 50 for each type of cost).
In the context of Ontario and Canada, our study's findings on decarbonization scenarios and their associated lifecycle costs hold significant relevance and implications. The region has been actively pursuing sustainable transportation initiatives and policies to combat climate change, reduce emissions, and promote greener mobility options. The government has shown commitment to sustainability through various policies and initiatives, including the Climate Change Action Plan and the GreenON Rebate Program. These efforts aim to incentivize the adoption of low-carbon and electric vehicles, thereby aligning with our research focus on Alternative Fuel Vehicles (AFVs) and their potential for emission reduction \citep{birchall2021climate}.

One of the key takeaways from our study is the importance of combining CAV penetration rates, eco-driving practices, and routing strategies to optimize emissions and travel time. Such findings are timely and relevant for Ontario, as the province has explored the potential of connected and automated vehicle technologies to improve traffic flow, reduce congestion, and improve transportation efficiency. In terms of public perception, our study highlights the need for cost reductions in Battery Electric Vehicles (BEVs) to enhance their competitiveness. This insight aligns with public sentiment towards electric vehicles, where concerns about higher upfront costs and charging infrastructure have been barriers to widespread adoption. 

\begin{figure}[h]
  \begin{center}
   \includegraphics[width=\textwidth]{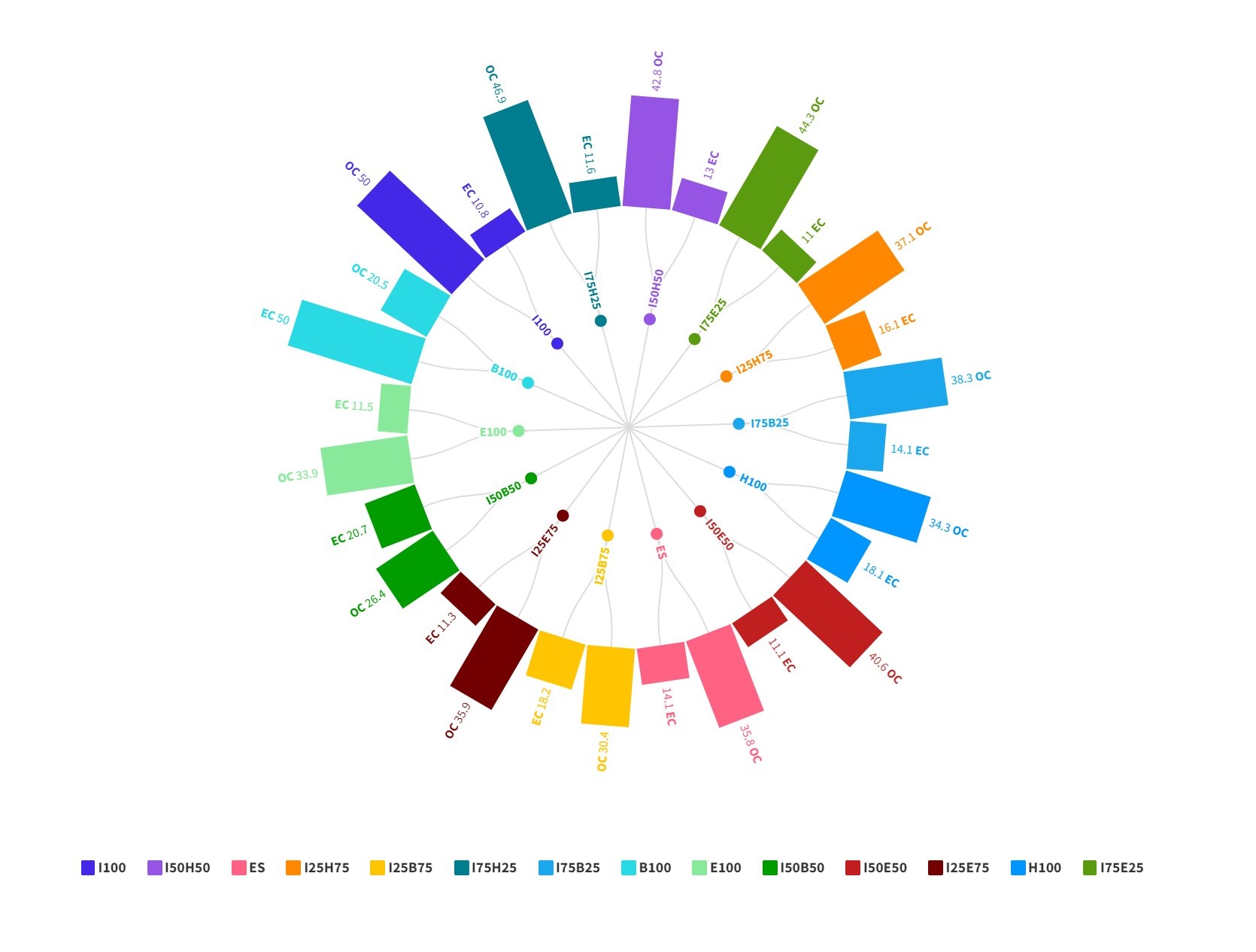}
   \caption{Cost Analysis using a transformed indicator}\label{costs}
  \end{center}
\end{figure}

It's important to note that combining the scores for each alternative and providing an overall score out of 100 would be misleading due to the significant difference in the scope of the costs. Therefore, we presented separate scores for emission-related costs and other-related costs to provide a clearer understanding of the relative rankings. By considering both emission and non-emission costs separately, this analysis allows us to assess the financial implications of different fuel options. It emphasizes the importance of cost reductions in BEVs to enhance their competitiveness in the market and highlights potential opportunities for HEVs and E-fuels as viable alternatives. This methodology helps us avoid potential misinterpretations by acknowledging the significant differences in cost scopes and providing a more nuanced understanding of the cost analysis results.
We have divided the costs in the unit of CAD per kilometer,  based on emission costs and other costs. 

{\section{Planning and Policy Implications}
\label{pol}
The implications of this study are far-reaching and can assist various stakeholders in making informed decisions to address environmental challenges and promote sustainable transportation. Here are the key implications based on the study's results and key findings:}

{\textbf{Promotion of Green Alternatives and Cost Reduction.} The study emphasizes the importance of Battery Electric Vehicles (BEVs) as a key decarbonization pathway with the least Well-to-Wheel (WTW) GHG emissions. However, Hybrid Electric Vehicles (HEVs) and e-fuels present more affordable and feasible short-term alternatives. To facilitate the adoption of these green technologies, governments should offer subsidies and promote cost-reduction strategies for BEVs, while continuing to support cleaner and more efficient battery technologies.}

{\textbf{Eco-Driving and Routing Integration for Efficiency.} The analysis demonstrates that eco-driving techniques consistently lead to reduced NOx emissions. Governments and policymakers should invest in promoting eco-driving education and awareness campaigns among the general public to foster environmentally responsible driving practices. Virtual Reality-based (VR) immersive simulators (e.g., \citet{ansar2023behavioural}) can play an important role in the training of drivers to adopt eco-driving. Learning eco-driving strategies can also be included in the process of acquiring and renewing a driver's license. The integration of Connected and Automated Vehicles (CAVs) with anticipatory routing has the potential to significantly reduce travel time and improve transportation efficiency. Governments, policymakers, and industry players should invest in CAV development and associated infrastructure to enhance the transportation network's effectiveness. \citet{FarooqBilalandDjavadian} has shown that E2ECAV based anticipatory routing can be particularly effective in congested areas in reducing travel times and increasing throughput. Such routing systems can be adopted by large cities, such as Toronto, as a part of their traffic management plan.}

{\textbf{Holistic and Multi-Modal Sustainable Transportation.} The study emphasizes the importance of adopting a comprehensive approach to sustainable transportation. Decision-makers should consider factors like fuel type combinations, routing strategies, and eco-driving practices together to achieve more sustainable and environmentally friendly transportation systems. While the focus of this study was only on vehicular traffic, in the long run, policymakers should also focus on creating integrated and multi-modal transportation systems that cater to different travel needs. Integrating various modes of transport, including public transit, cycling lanes, and pedestrian-friendly infrastructure, can reduce reliance on individual vehicles and promote sustainable mobility options.}

%\subsection{Dynamic Routing Systems} Implementing dynamic routing systems that can optimize travel routes in real-time based on current conditions and congestion can significantly improve travel time and overall transportation efficiency.

{\textbf{Continuous Policy Evaluation and Collaboration.} Governments and policymakers should continuously evaluate the effectiveness of existing policies and interventions. Applying optimization techniques to identify optimal combinations of policies, technologies, and behaviour can help achieve specific environmental objectives. The agent-based microsimulations, such as the one used in this study, can play a crucial role in such evaluation and optimization processes. To address the challenges of transportation decarbonization comprehensively, interdisciplinary collaboration between the transportation sector, energy sector, and urban planning is essential. Understanding the synergies and trade-offs among these sectors can lead to more effective and harmonized policies.}

{\textbf{Consumer Choices and Sustainable Policies.} The general public plays a crucial role in the transition to sustainable transportation.}
{The current study's findings underscore the importance of consumer choices in driving the transition to sustainable transportation. The cost analysis reveals that while BEVs provide significant GHG reductions, their higher initial costs make them less competitive without incentives. Conversely, HEVs and e-fuels offer a balance between cost efficiency and emissions reduction, appealing to consumers focused on affordability. Encouraging the adoption of these technologies, through eco-driving practices and supportive policies, can optimize emissions and reduce travel costs. The findings also highlight that aligning consumer preferences with greener technologies, backed by incentives, is crucial for achieving meaningful environmental gains.}

\section{Limitations}
\label{limits}
While the study provides valuable insights into the effects of various factors on greenhouse gas emissions, travel time, and associated costs in the context of smart and sustainable transportation, there are some limitations that should be considered. The analysis does not fully account for human behaviour and its influence on transportation choices. Factors such as public perception, social acceptance of new technologies, and behavioural changes in response to policies might be challenging to incorporate fully. In addition, the analysis is based on specific technological assumptions, such as the performance and availability of future technologies like CAVs or e-fuels. These assumptions can affect the realism of the scenarios considered. Policies and regulations related to transportation and environmental sustainability are subject to change over time. The study's static analysis may not capture the evolving nature of policies and their potential impacts on consumer preference. The study might not comprehensively analyze the socioeconomic implications of different decarbonization scenarios, such as income distribution, government subsidies, or social equity. While the study focuses on transportation, its environmental impacts are interconnected with other sectors like the energy sector as the leading emitter. The study might not fully capture the synergies or trade-offs between transportation policies and those of other sectors. {While significant mode shifts toward public transit and active transportation are important long-term goals, achieving such shifts may not be feasible in the short term due to infrastructure and behaviour change related limitations. Consequently, decarbonization options related to passenger vehicles, including electric vehicles, must be explored.} Acknowledging these limitations is crucial for interpreting the study's findings and informing future research and policy decision makers to achieve more robust and comprehensive solutions for smart and sustainable cities and society.

\section{Conclusions}
\label{con}
Our study's comprehensive simulation-based decarbonization scenarios analysis, coupled with the implementation of the Transformer neural network model, yields deeper insights into the specific effects of fuel type combinations, Connected and Automated Vehicles (CAVs) penetration rates, eco-driving practices, and routing strategies on Greenhouse Gas (GHG) emissions, Nitrogen Oxides (NOx) emissions, travel time, and associated costs. By incorporating the Transformer model, we significantly improve the accuracy of predictions for GHG emissions, NOx emissions, and average speed, particularly in handling long-term dependencies in time-series data. This advanced modelling approach enhances the reliability and precision of our results, enabling more robust assessments of the environmental impact of different scenarios.

In terms of GHG emissions, the findings demonstrate that 100\% BEVs have the lowest WTW GHG emissions due to the absence of tailpipe emissions. This highlights the potential of BEVs as a greener alternative. However, it is important to consider the current limitations of BEVs, such as high costs and limited production capacity, which make them less accessible to a wider consumer base. The study also highlights the significant role of other green alternatives, such as HEVs and e-fuels, in achieving emission reductions. The analysis of NOx emissions reveals that BEVs have considerably lower upstream emissions compared to other alternatives, positioning them as a promising solution for reducing NOx emissions. Additionally, the adoption of eco-driving techniques consistently resulted in reduced NOx emissions across different routing strategies. This emphasizes the importance of promoting eco-driving practices to further mitigate the environmental impact of transportation. When examining travel time, it is observed that the impact of fuel type on average travel time in the scenarios considered is negligible. However, the routing strategy emerged as a significant factor that influenced travel time. Myopic routing, which aims to minimize travel time based on current conditions, and anticipatory routing, which optimizes travel routes to avoid congestion, are found to have distinct impacts. Anticipatory routing, especially when combined with CAV integration, demonstrated the potential to significantly reduce travel time, providing a more efficient and optimized transportation network.

{The lifecycle cost analysis conducted in this study provides a comprehensive evaluation of both emission and non-emission costs across different fuel technologies and vehicle types, including Battery Electric Vehicles (BEVs), Hybrid Electric Vehicles (HEVs), Internal Combustion Engine Vehicles (ICEVs), and e-fuels, all within a 100\% Connected and Automated Vehicles (CAVs) scenario. BEVs offer the greatest reduction in greenhouse gas (GHG) emissions, with costs as low as \$0.08/km, compared to \$0.50/km for ICEVs. HEVs emerged as an intermediate solution with a GHG cost of \$0.30/km. In terms of NOx emissions, BEVs showed almost negligible levels (\$0.01/km), making them the most environmentally friendly option, while ICEVs incurred significantly higher NOx costs at \$0.12/km.
Non-emission costs, such as travel time, fuel, and operation and maintenance, also play a critical role in assessing the overall cost-effectiveness of each technology. With CAV integration and optimized routing strategies, travel time was reduced by up to 25\%, saving around \$0.08/km. Fuel costs for BEVs averaged \$0.10/km, lower than ICEVs at \$0.14/km, while HEVs presented a middle ground at \$0.12/km. BEVs also offered reduced operation and maintenance costs due to fewer mechanical components (\$0.05/km), compared to ICEVs (\$0.08/km) and HEVs (\$0.06/km). Despite these savings, the higher upfront cost of BEVs remains a barrier, though federal and provincial incentives in Canada help offset this. Our cost ranking, which assigns equal weights to emission and non-emission costs, shows that while BEVs score highest in emission reductions, ICEVs lead in non-emission costs, making them more affordable for consumers in the short term. HEVs and e-fuels balance emissions and economic efficiency.}

Overall, The integration of CAVs, combined with anticipatory routing and eco-driving practices, demonstrated the potential to achieve significant improvements in travel time and further enhance environmental sustainability.
These findings emphasize the importance of adopting a holistic approach to sustainable transportation. By considering these factors in detail, we can guide decision-making processes toward achieving more sustainable and environmentally friendly transportation systems.

\section*{Declaration of Competing Interest}
The authors declare that they have no known competing financial interests or personal relationships that could have appeared to influence the work reported in this paper.

\section*{Data Availability}
Data will be made available upon request.

\section*{Acknowledgements}
This research is funded by a grant from the Canada Research Chair program in Disruptive Transportation Technologies and Services (CRC-2021-00480) and NSERC Discovery (RGPIN-2020-04492) fund. We extend our sincere gratitude to Dr. Mehdi Meshkani and Dr. Lama Alfaseeh for their invaluable assistance and guidance in initiating the base simulation.




%The citation must be used in following style: \cite{article-minimal} \cite{article-full} \cite{article-crossref} \cite{whole-journal}.
%% References with BibTeX dat\cite{}e:\cite{}e
%\bibliographystyle{elsarticle-harv}
\bibliographystyle{elsarticle-harv}
\bibliography{reff}

\begin{thebibliography}{50}
\expandafter\ifx\csname natexlab\endcsname\relax\def\natexlab#1{#1}\fi
\providecommand{\url}[1]{\texttt{#1}}
\providecommand{\href}[2]{#2}
\providecommand{\path}[1]{#1}
\providecommand{\DOIprefix}{doi:}
\providecommand{\ArXivprefix}{arXiv:}
\providecommand{\URLprefix}{URL: }
\providecommand{\Pubmedprefix}{pmid:}
\providecommand{\doi}[1]{\href{http://dx.doi.org/#1}{\path{#1}}}
\providecommand{\Pubmed}[1]{\href{pmid:#1}{\path{#1}}}
\providecommand{\bibinfo}[2]{#2}
\ifx\xfnm\relax \def\xfnm[#1]{\unskip,\space#1}\fi
%Type = Misc
\bibitem[{Alexander-Kearns et~al.(2016)Alexander-Kearns, Peterson and Cassady}]{alexander2016impact}
\bibinfo{author}{Alexander-Kearns, M.}, \bibinfo{author}{Peterson, M.}, \bibinfo{author}{Cassady, A.}, \bibinfo{year}{2016}.
\newblock \bibinfo{title}{The impact of vehicle automation on carbon emissions}.
\newblock \bibinfo{howpublished}{\url{https://www.americanprogress.org/article/the-impact-of-vehicle-automation-on-carbon-emissions-where-uncertainty-lies/}}.
\newblock \bibinfo{note}{Accessed on 08/02/2023}.
%Type = Inproceedings
\bibitem[{Alfaseeh et~al.(2018)Alfaseeh, Djavadian and Farooq}]{alfaseeh2018impact}
\bibinfo{author}{Alfaseeh, L.}, \bibinfo{author}{Djavadian, S.}, \bibinfo{author}{Farooq, B.}, \bibinfo{year}{2018}.
\newblock \bibinfo{title}{Impact of distributed routing of intelligent vehicles on urban traffic}, in: \bibinfo{booktitle}{2018 IEEE International Smart Cities Conference (ISC2)}, \bibinfo{organization}{IEEE}. pp. \bibinfo{pages}{1--7}.
%Type = Inproceedings
\bibitem[{Alfaseeh et~al.(2019)Alfaseeh, Djavadian, Tu, Farooq and Hatzopoulou}]{alfaseeh2019multi}
\bibinfo{author}{Alfaseeh, L.}, \bibinfo{author}{Djavadian, S.}, \bibinfo{author}{Tu, R.}, \bibinfo{author}{Farooq, B.}, \bibinfo{author}{Hatzopoulou, M.}, \bibinfo{year}{2019}.
\newblock \bibinfo{title}{Multi-objective eco-routing in a distributed routing framework}, in: \bibinfo{booktitle}{2019 IEEE International Smart Cities Conference (ISC2)}, \bibinfo{organization}{IEEE}. pp. \bibinfo{pages}{747--752}.
%Type = Article
\bibitem[{Alfaseeh and Farooq(2020)}]{alfaseeh2020deep}
\bibinfo{author}{Alfaseeh, L.}, \bibinfo{author}{Farooq, B.}, \bibinfo{year}{2020}.
\newblock \bibinfo{title}{Deep learning based proactive multi-objective eco-routing strategies for connected and automated vehicles}.
\newblock \bibinfo{journal}{Frontiers in Future Transportation} \bibinfo{volume}{1}, \bibinfo{pages}{6}.
%Type = Article
\bibitem[{Alfaseeh et~al.(2020)Alfaseeh, Tu, Farooq and Hatzopoulou}]{alfaseeh2020greenhouse}
\bibinfo{author}{Alfaseeh, L.}, \bibinfo{author}{Tu, R.}, \bibinfo{author}{Farooq, B.}, \bibinfo{author}{Hatzopoulou, M.}, \bibinfo{year}{2020}.
\newblock \bibinfo{title}{Greenhouse gas emission prediction on road network using deep sequence learning}.
\newblock \bibinfo{journal}{Transportation Research Part D: Transport and Environment} \bibinfo{volume}{88}, \bibinfo{pages}{102593}.
%Type = Article
\bibitem[{Althor et~al.(2016)Althor, Watson and Fuller}]{althor2016global}
\bibinfo{author}{Althor, G.}, \bibinfo{author}{Watson, J.E.}, \bibinfo{author}{Fuller, R.A.}, \bibinfo{year}{2016}.
\newblock \bibinfo{title}{Global mismatch between greenhouse gas emissions and the burden of climate change}.
\newblock \bibinfo{journal}{Scientific reports} \bibinfo{volume}{6}, \bibinfo{pages}{1--6}.
%Type = Article
\bibitem[{Ansar et~al.(2023)Ansar, Alsaleh and Farooq}]{ansar2023behavioural}
\bibinfo{author}{Ansar, M.S.}, \bibinfo{author}{Alsaleh, N.}, \bibinfo{author}{Farooq, B.}, \bibinfo{year}{2023}.
\newblock \bibinfo{title}{Behavioural modelling of automated to manual control transition in conditionally automated driving}.
\newblock \bibinfo{journal}{Transportation research part F: traffic psychology and behaviour} \bibinfo{volume}{94}, \bibinfo{pages}{422--435}.
%Type = Article
\bibitem[{Axsen et~al.(2022)Axsen, Bhardwaj and Crawford}]{axsen2022comparing}
\bibinfo{author}{Axsen, J.}, \bibinfo{author}{Bhardwaj, C.}, \bibinfo{author}{Crawford, C.}, \bibinfo{year}{2022}.
\newblock \bibinfo{title}{Comparing policy pathways to achieve 100\% zero-emissions vehicle sales by 2035}.
\newblock \bibinfo{journal}{Transportation Research Part D: Transport and Environment} \bibinfo{volume}{112}, \bibinfo{pages}{103488}.
%Type = Inproceedings
\bibitem[{Barth et~al.(2011)Barth, Mandava, Boriboonsomsin and Xia}]{barth2011dynamic}
\bibinfo{author}{Barth, M.}, \bibinfo{author}{Mandava, S.}, \bibinfo{author}{Boriboonsomsin, K.}, \bibinfo{author}{Xia, H.}, \bibinfo{year}{2011}.
\newblock \bibinfo{title}{Dynamic eco-driving for arterial corridors}, in: \bibinfo{booktitle}{2011 IEEE forum on integrated and sustainable transportation systems}, \bibinfo{organization}{IEEE}. pp. \bibinfo{pages}{182--188}.
%Type = Article
\bibitem[{Bhardwaj et~al.(2020)Bhardwaj, Axsen, Kern and McCollum}]{bhardwaj2020have}
\bibinfo{author}{Bhardwaj, C.}, \bibinfo{author}{Axsen, J.}, \bibinfo{author}{Kern, F.}, \bibinfo{author}{McCollum, D.}, \bibinfo{year}{2020}.
\newblock \bibinfo{title}{Why have multiple climate policies for light-duty vehicles? policy mix rationales, interactions and research gaps}.
\newblock \bibinfo{journal}{Transportation Research Part A: Policy and Practice} \bibinfo{volume}{135}, \bibinfo{pages}{309--326}.
%Type = Article
\bibitem[{Bieker(2021)}]{bieker2021global}
\bibinfo{author}{Bieker, G.}, \bibinfo{year}{2021}.
\newblock \bibinfo{title}{A global comparison of the life-cycle greenhouse gas emissions of combustion engine and electric passenger cars}.
\newblock \bibinfo{journal}{communications} \bibinfo{volume}{49}, \bibinfo{pages}{847129--102}.
%Type = Article
\bibitem[{Birchall and Bonnett(2021)}]{birchall2021climate}
\bibinfo{author}{Birchall, S.J.}, \bibinfo{author}{Bonnett, N.}, \bibinfo{year}{2021}.
\newblock \bibinfo{title}{Climate change adaptation policy and practice: The role of agents, institutions and systems}.
\newblock \bibinfo{journal}{Cities} \bibinfo{volume}{108}, \bibinfo{pages}{103001}.
%Type = Article
\bibitem[{Coloma et~al.(2020)Coloma, Garcia, Boggio-Marzet and Monz{\'o}n}]{coloma2020developing}
\bibinfo{author}{Coloma, J.}, \bibinfo{author}{Garcia, M.}, \bibinfo{author}{Boggio-Marzet, A.}, \bibinfo{author}{Monz{\'o}n, A.}, \bibinfo{year}{2020}.
\newblock \bibinfo{title}{Developing eco-driving strategies considering city characteristics}.
\newblock \bibinfo{journal}{Journal of Advanced Transportation} \bibinfo{volume}{2020}.
%Type = Article
\bibitem[{Dai et~al.(2019)Dai, Kelly, Gaines and Wang}]{dai2019life}
\bibinfo{author}{Dai, Q.}, \bibinfo{author}{Kelly, J.C.}, \bibinfo{author}{Gaines, L.}, \bibinfo{author}{Wang, M.}, \bibinfo{year}{2019}.
\newblock \bibinfo{title}{Life cycle analysis of lithium-ion batteries for automotive applications}.
\newblock \bibinfo{journal}{Batteries} \bibinfo{volume}{5}, \bibinfo{pages}{48}.
%Type = Article
\bibitem[{Diaz(2020)}]{diaz2020electric}
\bibinfo{author}{Diaz, M.N.}, \bibinfo{year}{2020}.
\newblock \bibinfo{title}{Electric vehicles: A primer on technology and selected policy issues}.
\newblock \bibinfo{journal}{Congressional Research Service. Washington, DC, USA} .
%Type = Article
\bibitem[{Djavadian et~al.(2020)Djavadian, Tu, Farooq and Hatzopoulou}]{djavadian2020multi}
\bibinfo{author}{Djavadian, S.}, \bibinfo{author}{Tu, R.}, \bibinfo{author}{Farooq, B.}, \bibinfo{author}{Hatzopoulou, M.}, \bibinfo{year}{2020}.
\newblock \bibinfo{title}{Multi-objective eco-routing for dynamic control of connected and automated vehicles}.
\newblock \bibinfo{journal}{Transportation Research Part D: Transport and Environment} \bibinfo{volume}{87}, \bibinfo{pages}{102513}.
%Type = Techreport
\bibitem[{Elgowainy et~al.(2016)Elgowainy, Han, Ward, Joseck, Gohlke, Lindauer, Ramsden, Biddy, Alexander, Barnhart et~al.}]{elgowainy2016cradle}
\bibinfo{author}{Elgowainy, A.}, \bibinfo{author}{Han, J.}, \bibinfo{author}{Ward, J.}, \bibinfo{author}{Joseck, F.}, \bibinfo{author}{Gohlke, D.}, \bibinfo{author}{Lindauer, A.}, \bibinfo{author}{Ramsden, T.}, \bibinfo{author}{Biddy, M.}, \bibinfo{author}{Alexander, M.}, \bibinfo{author}{Barnhart, S.}, et~al., \bibinfo{year}{2016}.
\newblock \bibinfo{title}{Cradle-to-grave lifecycle analysis of US light duty vehicle-fuel pathways: a greenhouse gas emissions and economic assessment of current (2015) and future (2025-2030) technologies}.
\newblock \bibinfo{type}{Technical Report}. Argonne National Lab.(ANL), Argonne, IL (United States).
%Type = Misc
\bibitem[{Farooq and Djavadian(2019)}]{FarooqBilalandDjavadian}
\bibinfo{author}{Farooq, B.}, \bibinfo{author}{Djavadian, S.}, \bibinfo{year}{2019}.
\newblock \bibinfo{title}{{Distributed Traffic Management System with Dynamic End-to-End Routing}}.
\newblock \bibinfo{note}{U.S. Provisional Pat. Ser. No. 62/865,725}.
%Type = Article
\bibitem[{Greenblatt and Shaheen(2015)}]{greenblatt2015automated}
\bibinfo{author}{Greenblatt, J.B.}, \bibinfo{author}{Shaheen, S.}, \bibinfo{year}{2015}.
\newblock \bibinfo{title}{Automated vehicles, on-demand mobility, and environmental impacts}.
\newblock \bibinfo{journal}{Current sustainable/renewable energy reports} \bibinfo{volume}{2}, \bibinfo{pages}{74--81}.
%Type = Article
\bibitem[{Gryparis et~al.(2020)Gryparis, Papadopoulos, Leligou and Psomopoulos}]{gryparis2020electricity}
\bibinfo{author}{Gryparis, E.}, \bibinfo{author}{Papadopoulos, P.}, \bibinfo{author}{Leligou, H.C.}, \bibinfo{author}{Psomopoulos, C.S.}, \bibinfo{year}{2020}.
\newblock \bibinfo{title}{Electricity demand and carbon emission in power generation under high penetration of electric vehicles. a european union perspective}.
\newblock \bibinfo{journal}{Energy Reports} \bibinfo{volume}{6}, \bibinfo{pages}{475--486}.
%Type = Article
\bibitem[{Horvath et~al.(2018)Horvath, Fasihi and Breyer}]{horvath2018techno}
\bibinfo{author}{Horvath, S.}, \bibinfo{author}{Fasihi, M.}, \bibinfo{author}{Breyer, C.}, \bibinfo{year}{2018}.
\newblock \bibinfo{title}{Techno-economic analysis of a decarbonized shipping sector: Technology suggestions for a fleet in 2030 and 2040}.
\newblock \bibinfo{journal}{Energy Conversion and Management} \bibinfo{volume}{164}, \bibinfo{pages}{230--241}.
%Type = Article
\bibitem[{Jenn et~al.(2019)Jenn, Azevedo and Michalek}]{jenn2019alternative}
\bibinfo{author}{Jenn, A.}, \bibinfo{author}{Azevedo, I.L.}, \bibinfo{author}{Michalek, J.J.}, \bibinfo{year}{2019}.
\newblock \bibinfo{title}{Alternative-fuel-vehicle policy interactions increase us greenhouse gas emissions}.
\newblock \bibinfo{journal}{Transportation Research Part A: Policy and Practice} \bibinfo{volume}{124}, \bibinfo{pages}{396--407}.
%Type = Techreport
\bibitem[{Jensterle et~al.(2019)Jensterle, Narita, Piria, Samadi, Prantner, Crone, Siegemund, Kan, Matsumoto, Shibata et~al.}]{jensterle2019role}
\bibinfo{author}{Jensterle, M.}, \bibinfo{author}{Narita, J.}, \bibinfo{author}{Piria, R.}, \bibinfo{author}{Samadi, S.}, \bibinfo{author}{Prantner, M.}, \bibinfo{author}{Crone, K.}, \bibinfo{author}{Siegemund, S.}, \bibinfo{author}{Kan, S.}, \bibinfo{author}{Matsumoto, T.}, \bibinfo{author}{Shibata, Y.}, et~al., \bibinfo{year}{2019}.
\newblock \bibinfo{title}{The role of clean hydrogen in the future energy systems of Japan and Germany: an analysis of existing mid-century scenarios and an investigation of hydrogen supply chains}.
\newblock \bibinfo{type}{Technical Report}. Wuppertal Institut.
%Type = Article
\bibitem[{Kato et~al.(2016)Kato, Ando, Kondo, Suzuki, Matsuhashi and Kobayashi}]{kato2016ecodriving}
\bibinfo{author}{Kato, H.}, \bibinfo{author}{Ando, R.}, \bibinfo{author}{Kondo, Y.}, \bibinfo{author}{Suzuki, T.}, \bibinfo{author}{Matsuhashi, K.}, \bibinfo{author}{Kobayashi, S.}, \bibinfo{year}{2016}.
\newblock \bibinfo{title}{The eco-driving effect of electric vehicles compared to conventional gasoline vehicles}.
\newblock \bibinfo{journal}{AIMS Energy} \bibinfo{volume}{4}, \bibinfo{pages}{804--816}.
\newblock \URLprefix \url{http://www.aimspress.com/journal/energy}, \DOIprefix\doi{10.3934/energy.2016.6.804}. \bibinfo{note}{received: 30 May 2016; Accepted: 08 October 2016; Published: 14 October 2016}.
%Type = Techreport
\bibitem[{Kelly et~al.(2022)Kelly, Elgowainy, Isaac, Ward, Islam, Rousseau, Sutherland, Wallington, Alexander, Muratori et~al.}]{kelly2022cradle}
\bibinfo{author}{Kelly, J.C.}, \bibinfo{author}{Elgowainy, A.}, \bibinfo{author}{Isaac, R.}, \bibinfo{author}{Ward, J.}, \bibinfo{author}{Islam, E.}, \bibinfo{author}{Rousseau, A.}, \bibinfo{author}{Sutherland, I.}, \bibinfo{author}{Wallington, T.J.}, \bibinfo{author}{Alexander, M.}, \bibinfo{author}{Muratori, M.}, et~al., \bibinfo{year}{2022}.
\newblock \bibinfo{title}{Cradle-to-Grave Lifecycle Analysis of US Light-Duty Vehicle-Fuel Pathways: A Greenhouse Gas Emissions and Economic Assessment of Current (2020) and Future (2030-2035) Technologies}.
\newblock \bibinfo{type}{Technical Report}. Argonne National Lab.(ANL), Argonne, IL (United States).
%Type = Article
\bibitem[{Liu(2015)}]{liu2015more}
\bibinfo{author}{Liu, X.}, \bibinfo{year}{2015}.
\newblock \bibinfo{title}{A more accurate method using moves (motor vehicle emission simulator) to estimate emission burden for regional-level analysis}.
\newblock \bibinfo{journal}{Journal of the Air \& Waste Management Association} \bibinfo{volume}{65}, \bibinfo{pages}{837--843}.
%Type = Article
\bibitem[{Logan et~al.(2021)Logan, Nelson, Brand and Hastings}]{logan2021phasing}
\bibinfo{author}{Logan, K.G.}, \bibinfo{author}{Nelson, J.D.}, \bibinfo{author}{Brand, C.}, \bibinfo{author}{Hastings, A.}, \bibinfo{year}{2021}.
\newblock \bibinfo{title}{Phasing in electric vehicles: Does policy focusing on operating emission achieve net zero emissions reduction objectives?}
\newblock \bibinfo{journal}{Transportation Research Part A: Policy and Practice} \bibinfo{volume}{152}, \bibinfo{pages}{100--114}.
%Type = Techreport
\bibitem[{McCubbin and Delucchi(1996)}]{mccubbin1996social}
\bibinfo{author}{McCubbin, D.R.}, \bibinfo{author}{Delucchi, M.A.}, \bibinfo{year}{1996}.
\newblock \bibinfo{title}{The social cost of the health effects of motor-vehicle air pollution}.
\newblock \bibinfo{type}{Technical Report}. University of California Transportation Center.
%Type = Article
\bibitem[{Naeem et~al.(2023)Naeem, Bhatti, Butt and Ahmed}]{naeem2023energy}
\bibinfo{author}{Naeem, H.M.Y.}, \bibinfo{author}{Bhatti, A.I.}, \bibinfo{author}{Butt, Y.A.}, \bibinfo{author}{Ahmed, Q.}, \bibinfo{year}{2023}.
\newblock \bibinfo{title}{Energy economization using connectivity-based eco-routing and driving for fleet of battery electric vehicles}.
\newblock \bibinfo{journal}{IEEE Transactions on Transportation Electrification} \bibinfo{volume}{10}, \bibinfo{pages}{1923--1934}.
%Type = Article
\bibitem[{Rahman and Thill(2023)}]{rahman2023impacts}
\bibinfo{author}{Rahman, M.M.}, \bibinfo{author}{Thill, J.C.}, \bibinfo{year}{2023}.
\newblock \bibinfo{title}{Impacts of connected and autonomous vehicles on urban transportation and environment: A comprehensive review}.
\newblock \bibinfo{journal}{Sustainable Cities and Society} , \bibinfo{pages}{104649}.
%Type = Article
\bibitem[{Rezaei et~al.(2019)Rezaei, Ebrahimnejad, Moosavi and Nikfarjam}]{rezaei2019green}
\bibinfo{author}{Rezaei, N.}, \bibinfo{author}{Ebrahimnejad, S.}, \bibinfo{author}{Moosavi, A.}, \bibinfo{author}{Nikfarjam, A.}, \bibinfo{year}{2019}.
\newblock \bibinfo{title}{A green vehicle routing problem with time windows considering the heterogeneous fleet of vehicles: two metaheuristic algorithms}.
\newblock \bibinfo{journal}{European Journal of Industrial Engineering} \bibinfo{volume}{13}, \bibinfo{pages}{507--535}.
%Type = Article
\bibitem[{Rony et~al.(2023)Rony, Mofijur, Hasan, Rasul, Jahirul, Ahmed, Kalam, Badruddin, Khan and Show}]{rony2023alternative}
\bibinfo{author}{Rony, Z.I.}, \bibinfo{author}{Mofijur, M.}, \bibinfo{author}{Hasan, M.}, \bibinfo{author}{Rasul, M.}, \bibinfo{author}{Jahirul, M.}, \bibinfo{author}{Ahmed, S.F.}, \bibinfo{author}{Kalam, M.}, \bibinfo{author}{Badruddin, I.A.}, \bibinfo{author}{Khan, T.Y.}, \bibinfo{author}{Show, P.L.}, \bibinfo{year}{2023}.
\newblock \bibinfo{title}{Alternative fuels to reduce greenhouse gas emissions from marine transport and promote un sustainable development goals}.
\newblock \bibinfo{journal}{Fuel} \bibinfo{volume}{338}, \bibinfo{pages}{127220}.
%Type = Techreport
\bibitem[{Rossi et~al.(2022)Rossi, Bargende, Kulzer, Chiodi, Massoud, Shrestha and Mauss}]{rossi2022analysis}
\bibinfo{author}{Rossi, E.}, \bibinfo{author}{Bargende, M.}, \bibinfo{author}{Kulzer, A.C.}, \bibinfo{author}{Chiodi, M.}, \bibinfo{author}{Massoud, E.}, \bibinfo{author}{Shrestha, K.}, \bibinfo{author}{Mauss, F.}, \bibinfo{year}{2022}.
\newblock \bibinfo{title}{Analysis of the Applicability of Water Injection in Combination with an eFuel for Knock Mitigation and Improved Engine Efficiency}.
\newblock \bibinfo{type}{Technical Report}. SAE Technical Paper.
%Type = Incollection
\bibitem[{Sabet and Farooq(2023)}]{RePEc:spr:sprchp:978-3-030-97940-9_126}
\bibinfo{author}{Sabet, S.}, \bibinfo{author}{Farooq, B.}, \bibinfo{year}{2023}.
\newblock \bibinfo{title}{{Energy-Smart Transportation Systems}}, in: \bibinfo{editor}{Fathi, M.}, \bibinfo{editor}{Zio, E.}, \bibinfo{editor}{Pardalos, P.M.} (Eds.), \bibinfo{booktitle}{{Handbook of Smart Energy Systems}}. \bibinfo{publisher}{Springer}. Springer Books, pp. \bibinfo{pages}{2003--2023}.
\newblock \URLprefix \url{https://ideas.repec.org/h/spr/sprchp/978-3-030-97940-9_126.html}, \DOIprefix\doi{10.1007/978-3-030-97940-9}.
%Type = Article
\bibitem[{Samaranayake et~al.(2024)Samaranayake, Chand, Sinha and Dixit}]{samaranayake2024impact}
\bibinfo{author}{Samaranayake, S.}, \bibinfo{author}{Chand, S.}, \bibinfo{author}{Sinha, A.}, \bibinfo{author}{Dixit, V.}, \bibinfo{year}{2024}.
\newblock \bibinfo{title}{Impact of connected and automated vehicles on the travel time reliability of an urban network}.
\newblock \bibinfo{journal}{International journal of transportation science and technology} \bibinfo{volume}{13}, \bibinfo{pages}{171--185}.
%Type = Article
\bibitem[{Shen et~al.(2020)Shen, Karbowski and Rousseau}]{shen2020minimum}
\bibinfo{author}{Shen, D.}, \bibinfo{author}{Karbowski, D.}, \bibinfo{author}{Rousseau, A.}, \bibinfo{year}{2020}.
\newblock \bibinfo{title}{A minimum principle-based algorithm for energy-efficient eco-driving of electric vehicles in various traffic and road conditions}.
\newblock \bibinfo{journal}{IEEE Transactions on Intelligent Vehicles} \bibinfo{volume}{5}, \bibinfo{pages}{725--737}.
%Type = Misc
\bibitem[{StatisticsCanada(2022)}]{CanadaStatistics}
\bibinfo{author}{StatisticsCanada}, \bibinfo{year}{2022}.
\newblock \bibinfo{title}{Automotive statistics}.
\newblock \bibinfo{howpublished}{\url{https://www.statcan.gc.ca/en/topics-start/automotive}}.
\newblock \bibinfo{note}{Accessed: July 07, 2023}.
%Type = Article
\bibitem[{Stringer and Joanis(2022)}]{stringer2022assessing}
\bibinfo{author}{Stringer, T.}, \bibinfo{author}{Joanis, M.}, \bibinfo{year}{2022}.
\newblock \bibinfo{title}{Assessing energy transition costs: Sub-national challenges in canada}.
\newblock \bibinfo{journal}{Energy Policy} \bibinfo{volume}{164}, \bibinfo{pages}{112879}.
%Type = Article
\bibitem[{Thierer and Watney(2016)}]{thierer2016comment}
\bibinfo{author}{Thierer, A.D.}, \bibinfo{author}{Watney, C.}, \bibinfo{year}{2016}.
\newblock \bibinfo{title}{Comment on the federal automated vehicles policy}.
\newblock \bibinfo{journal}{Available at SSRN 2876832} .
%Type = Article
\bibitem[{Treiber et~al.(2000)Treiber, Hennecke and Helbing}]{treiber2000congested}
\bibinfo{author}{Treiber, M.}, \bibinfo{author}{Hennecke, A.}, \bibinfo{author}{Helbing, D.}, \bibinfo{year}{2000}.
\newblock \bibinfo{title}{Congested traffic states in empirical observations and microscopic simulations}.
\newblock \bibinfo{journal}{Physical review E} \bibinfo{volume}{62}, \bibinfo{pages}{1805}.
%Type = Article
\bibitem[{Tu et~al.(2019)Tu, Alfaseeh, Djavadian, Farooq and Hatzopoulou}]{tu2019quantifying}
\bibinfo{author}{Tu, R.}, \bibinfo{author}{Alfaseeh, L.}, \bibinfo{author}{Djavadian, S.}, \bibinfo{author}{Farooq, B.}, \bibinfo{author}{Hatzopoulou, M.}, \bibinfo{year}{2019}.
\newblock \bibinfo{title}{Quantifying the impacts of dynamic control in connected and automated vehicles on greenhouse gas emissions and urban no2 concentrations}.
\newblock \bibinfo{journal}{Transportation Research Part D: Transport and Environment} \bibinfo{volume}{73}, \bibinfo{pages}{142--151}.
%Type = Article
\bibitem[{Tu et~al.(2020)Tu, Gai, Farooq, Posen and Hatzopoulou}]{tu2020electric}
\bibinfo{author}{Tu, R.}, \bibinfo{author}{Gai, Y.J.}, \bibinfo{author}{Farooq, B.}, \bibinfo{author}{Posen, D.}, \bibinfo{author}{Hatzopoulou, M.}, \bibinfo{year}{2020}.
\newblock \bibinfo{title}{Electric vehicle charging optimization to minimize marginal greenhouse gas emissions from power generation}.
\newblock \bibinfo{journal}{Applied Energy} \bibinfo{volume}{277}, \bibinfo{pages}{115517}.
%Type = Article
\bibitem[{Tzeiranaki et~al.(2023)Tzeiranaki, Economidou, Bertoldi, Thiel, Fontaras, Clementi and De~Los~Rios}]{tzeiranaki2023impact}
\bibinfo{author}{Tzeiranaki, S.T.}, \bibinfo{author}{Economidou, M.}, \bibinfo{author}{Bertoldi, P.}, \bibinfo{author}{Thiel, C.}, \bibinfo{author}{Fontaras, G.}, \bibinfo{author}{Clementi, E.L.}, \bibinfo{author}{De~Los~Rios, C.F.}, \bibinfo{year}{2023}.
\newblock \bibinfo{title}{The impact of energy efficiency and decarbonisation policies on the european road transport sector}.
\newblock \bibinfo{journal}{Transportation Research Part A: Policy and Practice} \bibinfo{volume}{170}, \bibinfo{pages}{103623}.
%Type = Article
\bibitem[{Ueckerdt et~al.(2021)Ueckerdt, Bauer, Dirnaichner, Everall, Sacchi and Luderer}]{ueckerdt2021potential}
\bibinfo{author}{Ueckerdt, F.}, \bibinfo{author}{Bauer, C.}, \bibinfo{author}{Dirnaichner, A.}, \bibinfo{author}{Everall, J.}, \bibinfo{author}{Sacchi, R.}, \bibinfo{author}{Luderer, G.}, \bibinfo{year}{2021}.
\newblock \bibinfo{title}{Potential and risks of hydrogen-based e-fuels in climate change mitigation}.
\newblock \bibinfo{journal}{Nature Climate Change} \bibinfo{volume}{11}, \bibinfo{pages}{384--393}.
%Type = Misc
\bibitem[{U.S.E.P.A.(2020)}]{United}
\bibinfo{author}{U.S.E.P.A.}, \bibinfo{year}{2020}.
\newblock \bibinfo{title}{Greenhouse gases equivalencies calculator - calculations and references}.
\newblock \bibinfo{howpublished}{\url{https://www.epa.gov/energy/greenhouse-gases-equivalencies-calculator-calculations-and-references}}.
\newblock \bibinfo{note}{Accessed: July 23, 2023}.
%Type = Article
\bibitem[{Veza et~al.(2023)Veza, Asy'ari, Idris, Epin, Fattah and Spraggon}]{veza2023electric}
\bibinfo{author}{Veza, I.}, \bibinfo{author}{Asy'ari, M.Z.}, \bibinfo{author}{Idris, M.}, \bibinfo{author}{Epin, V.}, \bibinfo{author}{Fattah, I.R.}, \bibinfo{author}{Spraggon, M.}, \bibinfo{year}{2023}.
\newblock \bibinfo{title}{Electric vehicle (ev) and driving towards sustainability: Comparison between ev, hev, phev, and ice vehicles to achieve net zero emissions by 2050 from ev}.
\newblock \bibinfo{journal}{Alexandria Engineering Journal} \bibinfo{volume}{82}, \bibinfo{pages}{459--467}.
%Type = Article
\bibitem[{Wadud et~al.(2016)Wadud, MacKenzie and Leiby}]{wadud2016help}
\bibinfo{author}{Wadud, Z.}, \bibinfo{author}{MacKenzie, D.}, \bibinfo{author}{Leiby, P.}, \bibinfo{year}{2016}.
\newblock \bibinfo{title}{Help or hindrance? the travel, energy and carbon impacts of highly automated vehicles}.
\newblock \bibinfo{journal}{Transportation Research Part A: Policy and Practice} \bibinfo{volume}{86}, \bibinfo{pages}{1--18}.
%Type = Article
\bibitem[{Wu et~al.(2018)Wu, Wang, Zheng, Sun, Zhao and Wang}]{wu2018life}
\bibinfo{author}{Wu, Z.}, \bibinfo{author}{Wang, M.}, \bibinfo{author}{Zheng, J.}, \bibinfo{author}{Sun, X.}, \bibinfo{author}{Zhao, M.}, \bibinfo{author}{Wang, X.}, \bibinfo{year}{2018}.
\newblock \bibinfo{title}{Life cycle greenhouse gas emission reduction potential of battery electric vehicle}.
\newblock \bibinfo{journal}{Journal of Cleaner Production} \bibinfo{volume}{190}, \bibinfo{pages}{462--470}.
%Type = Article
\bibitem[{Xu et~al.(2021)Xu, Li, Liu and Zhao}]{xu2021overview}
\bibinfo{author}{Xu, N.}, \bibinfo{author}{Li, X.}, \bibinfo{author}{Liu, Q.}, \bibinfo{author}{Zhao, D.}, \bibinfo{year}{2021}.
\newblock \bibinfo{title}{An overview of eco-driving theory, capability evaluation, and training applications}.
\newblock \bibinfo{journal}{Sensors} \bibinfo{volume}{21}, \bibinfo{pages}{6547}.
%Type = Article
\bibitem[{Zavalko(2018)}]{zavalko2018applying}
\bibinfo{author}{Zavalko, A.}, \bibinfo{year}{2018}.
\newblock \bibinfo{title}{Applying energy approach in the evaluation of eco-driving skill and eco-driving training of truck drivers}.
\newblock \bibinfo{journal}{Transportation Research Part D: Transport and Environment} \bibinfo{volume}{62}, \bibinfo{pages}{672--684}.

\end{thebibliography}



%\clearpage

%%%% This page is for instructions only, once the article is finalize please omit the below text before creating the final PDF
%\normalMode

\end{document}